%
%
%

\input epsf
%
  \MAINTITLE={Metallicity indicators across the spectrum of composite 
  \newline stellar populations}      
%
%
%
  \SUBTITLE={ ????? }       
  \AUTHOR={ C. S.~M\"oller, U. Fritze - v. Alvensleben, 
   K. J. Fricke }
%
  \INSTITUTE={ Universit\"atssternwarte, Geismarlandstr. 11, D - 37083
  G\"ottingen, Germany}
%
  \OFFPRINTS={ Claudia S. M\"oller}
  \DATE={Received 28 December 1995, accepted 10 June 1996} 
%
  \ABSTRACT={ With a chemically consistent evolutionary synthesis approach we 
  follow the enrichment of individual chemical elements in the ISM. We 
  describe the time evolution of broad band colors U to K and 
  metal indices (Fe5335, Fe5270, Mg$_2$, Mgb), taking into account 
  the increasing initial metallicities of successive generations
  of stars. 
  Using specific star formation prescriptions for the different spectral 
  types of galaxies, we show our model galaxy broad 
  band colors, synthetic spectra, and ISM abundances 
  are in agreement with observations of 
  nearby samples and template galaxies of the respective spectral types.
  
  We compare our synthetic Mg$_2$ indices with observations and 
  study the time
  evolution of the stellar index [MgFe] for composite elliptical galaxy 
  models and for simple stellar populations
  of various metallicities.
  We show how the luminosity weighted 
  stellar metallicity indicators in different wavelength bands from U to K 
  compare to each other for the various 
  star formation histories and how they evolve in time.
  We analyse as a function of wavelength the luminosity contributions of
  stellar subpopulations with different metallicities in the various 
  galaxy types.  
  The relations between luminosity weighted stellar metallicity indicators in 
  different passbands, stellar absorption indices and 
  ISM abundances are shown to depend significantly on the star formation history 
  and evolutionary stage.}  

  \KEYWORDS={ Galaxies: abundances; evolution; stellar content;
elliptical; spiral.}
  \THESAURUS={ 11.01.1, 11.05.2, 11.19.5, 11.05.1, 11.19.2}
\maketitle
\MAINTITLERUNNINGHEAD{Metallicity indicators across the spectrum
of composite stellar populations}
\AUTHORRUNNINGHEAD{C. S. M\"oller et al. }
\titlea{Introduction}

Strong efforts have been devoted to calibrate and intercompare stellar 
metallicity indicators determined for galaxies in various wavelength regimes,
e.g. in the NIR where the indices are
less age dependent than in the optical wavelength range.
Many of the problems arising in this context have to do with the fact that
galaxies are complex stellar systems containing subpopulations of various ages
and metallicities. The relative light contributions of these different
subpopulations can not only depend on the star formation (SF) 
history of the galaxy 
but also on the wavelength range of interest.
The majority of stars in galaxies have metallicities about solar or
lower. Only in the
bulges of spirals and in the central regions of giant ellipticals 
may subpopulations of stars with metal abundances
reaching a few times solar be present. 
Arimoto \& Yoshii (1986) presented
a first attempt to explore the effect of the continous enrichment of the
successive stellar generations on the photometric evolution of composite 
stellar populations. 
Our chemically consistent model follows both
the spectral evolution of successive stellar generations, which may have
different metallicities, and the 
chemical enrichment of the ISM.

Another well-known problem is the age-metallicity-degeneracy
(e.g. O'Connell 1976, Bica \& Alloin 1986, Bica et al. 1988, Faber et al. 1995). 
A lot of effort has been made to disentangle this degeneracy by 
measuring stellar metal indices
in different stellar systems (Burstein et al. 1984, Bica \& Alloin 1987 ff), 
and recently for a set of individual stars covering a wide range in metallicity,
effective temperature and surface gravity (Gorgas et al. 1993).
Bica (1988), Worthey (1994), Worthey et al. (1994), Bressan et al. (1995)
use these varous libraries of metal indices for modelling the indices of 
composite stellar populations.
Until now, calibrations of theoretical indices were taken mainly from single burst
populations or star clusters. 

In this paper we show the time evolution of some indices commonly used
for galaxies with different star formation histories taking into
account in a consistent way, 
the evolution of the initial metallicity for successive stellar generations.

The present paper is organized as follows: in Sec. 2 we describe our chemically
consistent evolutionary code and in Sec. 3 we present the comparison of our results 
with observational data. In Sec. 4 we analyse the wavelength dependence of the 
luminosity
weighted stellar metallicities and in Sec. 5 we discuss
the relations between stellar indices
and gas metallicity for galaxies of various spectral types. 
Sec. 6 summarizes our conclusions.

\titlea{The Evolutionary Synthesis Model}

\titleb{Computational method}

Our model is basically a synthesis of a photometric evolutionary code, similar
to Bruzual \& Charlot (1993), and a chemical evolution scheme for ISM 
abundances described by Matteucci et al. (1991 ff).

The initial condition is a homogeneous gas cloud of a given mass and 
primordial metallicity $Z= 10^{-8}$ (closed box 1 - zone model,
instantaneous mixing).
The total stellar mass, gas content and abundances are calculated by integrating
Tinsley's equations (e.g. Tinsley 1972). Supernova I contributions to [Fe/H] 
from carbon deflation of white dwarfs in binary systems
are taken into account in the way described by Matteucci \& Tornamb\`e (1987)
and Matteucci et al. (1991). 

Our model follows the chemical enrichment of 
a series of individual elements in
the ISM as well as the metallicity evolution
of the stellar population, together with its spectral evolution.
The evolution of each star
is followed in the HR diagram from birth to its final phases
so that at each timestep the distribution of all stars over the
HRD is known (Fritze - v. Alvensleben 1989, Kr\"uger et al. 1995).
This HRD population is used to compute the integrated colors in two 
different ways.
At any given timestep an integrated galaxy spectrum is synthesized 
from our library of stellar 
spectra. This library is chosen from stellar atmosphere models (Kurucz 1992)
for the five metallicities [Fe/H]= -2.5, -1.5, -0.5, -0.3, 0.3.
With tabulated filter characteristics for UBVRIJHKL bands,
spectrophotometric colors can be deduced from the synthetic galaxy
spectrum (see also Guiderdoni \& Rocca - Volmerange 1987).
We also derive the integrated colors of 
the galaxy population from photometric calibrations (UBVRI: Green et al. 
1987, V-K: Taylor et al. 1987) as a function of effective temperature $T_{eff}$ and 
luminosity class $LC$ for all stellar evolutionary tracks.
The agreement
between both methods is better than 0.1 mag if emission lines from the ISM are
excluded. 

In the {\bf chemically consistent model} the evolution of the HRD 
population is followed with the five sets of stellar tracks for five different
metallicities 
($Z= 10^{-4}, 10^{-3}, 4 \times 10^{-3}, 10^{-2}, 4 \times 10^{-2}$)
compiled by Einsel et al.  (1995).
We distinguish five discrete metallicity ranges, one for each set of
stellar tracks. If the ISM metallicity increases above one of our
limiting metallicities, the evolution of stars formed thereafter are
followed with the tracks for the higher metallicity.
The HRD population is calculated for each met\-al\-li\-ci\-ty  
together with the
appropriate photometric color calibration. 
The total galaxy luminosities (U to K) at any given timestep are obtained 
by coadding the various single metallicity contributions.
Synthesizing the galaxy spectra is done by summing the stellar 
spectra, weighted 
by the distribution of the HRD population. 

Our computational code follows the distribution of stellar populations over 
the five discrete metallicity re\-gimes and calculates their respective luminosity 
contributions
in specific wavelength regimes (M\"oller 1995).

\titleb{Input physics}

We use detailed stellar evolutionary tracks for the five metallicities
$Z= 10^{-4}, 10^{-3}, 4 \times 10^{-3}, 10^{-2}, 4 \times 10^{-2}$ to follow the
metal enrichment of the stellar population. These tracks include all important
stages from the main sequence to core helium exhaustion or AGB (onset of
thermal pulses) and were 
calculated from various authors. These sets of stellar tracks were compiled and
first used by Einsel (1992), who devoted much effort in selecting from the
vast literature those tracks, which are as consistent as possible 
with the physical assumptions. 
These sets altogether comprise 170 stellar tracks with 2882 HRD - states.
Stars with metallicities higher than twice solar are not included in our model,
however
their contribution to the global spectral properties of normal galaxies is
negligible.
The colors have been calibrated with Green et al. (1987) color tables. Linear
interpolation has been done in the physical variables log g, T$_{eff}$, and Z
of these tables, to attach UBVRI-colors to each of the discrete points of our
stellar evolutionary tracks in the HRD. The temperature range of the tables
reaches up to 20 000 $K$, above which colors fortunately become almost metallicity
independent. (V-K) has been calibrated with Bessel \& Brett (1988) tables,
excepted for the red giants where we use the table by Taylor et al. (1987). 
Einsel et al. (1995) 
give a detailed description of these tracks and the sources 
from which they were taken. 

The basic parameters for this type of evolutionary model are the initial 
mass function (IMF)
and the SF law.
The IMF is taken from Scalo (1986) in the form $\phi (m) \sim m^{1+x}$ 
with indices x=0.25, 1.35 and 1.70 between the limits
0.15M$_\odot $, 1M$_\odot $, 2M$_\odot $, 60M$_\odot $, respectively. 
We take into account the
dark matter in normalizing the IMF to the fraction of visible mass FVM = 0.5 
(Bahcall et al. 1992).

Following Sandage (1986) 
we describe the various spectrophotometric galaxy types with specific star 
formation rates (cf. Table 1). Besides these galaxy types (=Hubble types), 
we calculate 
simple stellar populations (SSPs) for different metallicities which form stars only
in an initial burst within the first $3 \cdot 10^8$ yr. The SSPs are a useful
description of star clusters and any SF history of a galaxy can be 
described in terms of a series of SSPs. This is already shown by Bica (1988), who 
takes into account for the first time, the full spread in metallicity.

\begtabfull
\tabcap{1}{Star formation rates $\psi (t)$ and 
    characteristic timescales for star formation t$_*$. $g:= {{M_G}\over{M_{tot}}}$
    for various galaxy types.}
\halign{#\hfil&\hfil#&\hfil#&\hfil#
&\quad\hfil#&\hfil#&\hfil#\cr
\noalign{\medskip}
 ~Type ~ & ~SFR(t) [M$_\odot$/ yr] ~                   & ~$t_*$ [Gyr]  \cr
\noalign{\medskip\hrule\medskip}
 E       & $1.2\cdot 10^{-9} M_{tot} \cdot e^{-t/t_*}$ & 1.0  \cr
 S0      & $4\cdot 10^{-10} g $                        & 2.2  \cr
 Sa      & $3\cdot 10^{-10} g $                        & 3.0  \cr
 Sb      & $2\cdot 10^{-10} g $                        & 5.0  \cr
 Sc      & $1\cdot 10^{-10} g $                        & 9.0  \cr
 Sd      & $0.6\cdot 10^{-10} M_{tot}$                 & 10.5 \cr
\noalign{\medskip\hrule}}
\endtab
    
The calculation of stellar lifetimes, yields and remnants are 
described in detail in Fritze - v. Alvensleben \& Gerhard (1994).
A detailed description of the chemical 
evolution model for individual ISM element abundances (e.g. $^{24}$Mg, 
$^{56}$Fe) which we use for our index 
calibrations are also given there.

We assign the 
stellar indices $Mg_2$, Mgb, Fe5335 and Fe5270 to all 
evolutionary stages of the five sets of stellar tracks using the 
empirical fitting functions of Gorgas et al.  (1993). 
For each stellar population we synthesize the indices from the individual
values of all their stars, weighting by their respective luminosity
at the wavelength of the index.
We also analyse the
[MgFe] (Bressan et al.  1995) 
index, which is the geometric mean of the two indices Mgb and 
$\langle Fe \rangle $,
[MgFe] = (Mgb $\langle Fe \rangle )^{0.5}$ where the index
$\langle Fe \rangle $ is the mean of the
indices Fe5270 and Fe5335 ($\langle Fe \rangle$ = (Fe5270 + Fe5335)/2). 
The global indices for model galaxies and SSPs at any time
are calculated by summing the stellar indices
for each stellar evolutionary stage, weighted by the total light 
contribution in the V band of all the stars at this position in the HRD.

The galaxy spectra are computed by a superposition of stellar atmosphere 
spectra from Kurucz (1992) weighted by the number of stars present in
each spectral type and luminosity class. We use a library with 264
stellar spectra altogether, 44 for each metallicity, covering the range 3500 K to
47500 K, which corresponds to spectral types from M3 to O5, and log g = 0.5 to 5.0.
For $T_{eff} \le 3500$K we supplement this library with observed spectra
(M4V, M6V) of solar metallicity
from the UV (Wu et al. 1983) to the optical (Gunn \& Stryker 1983)
and adding a black body spectra in the IR. 

\titlea{Model Galaxies and Comparison with Observation}

\begfig 18 cm 
\vskip -18 cm
\epsfysize=6 cm
{\epsffile{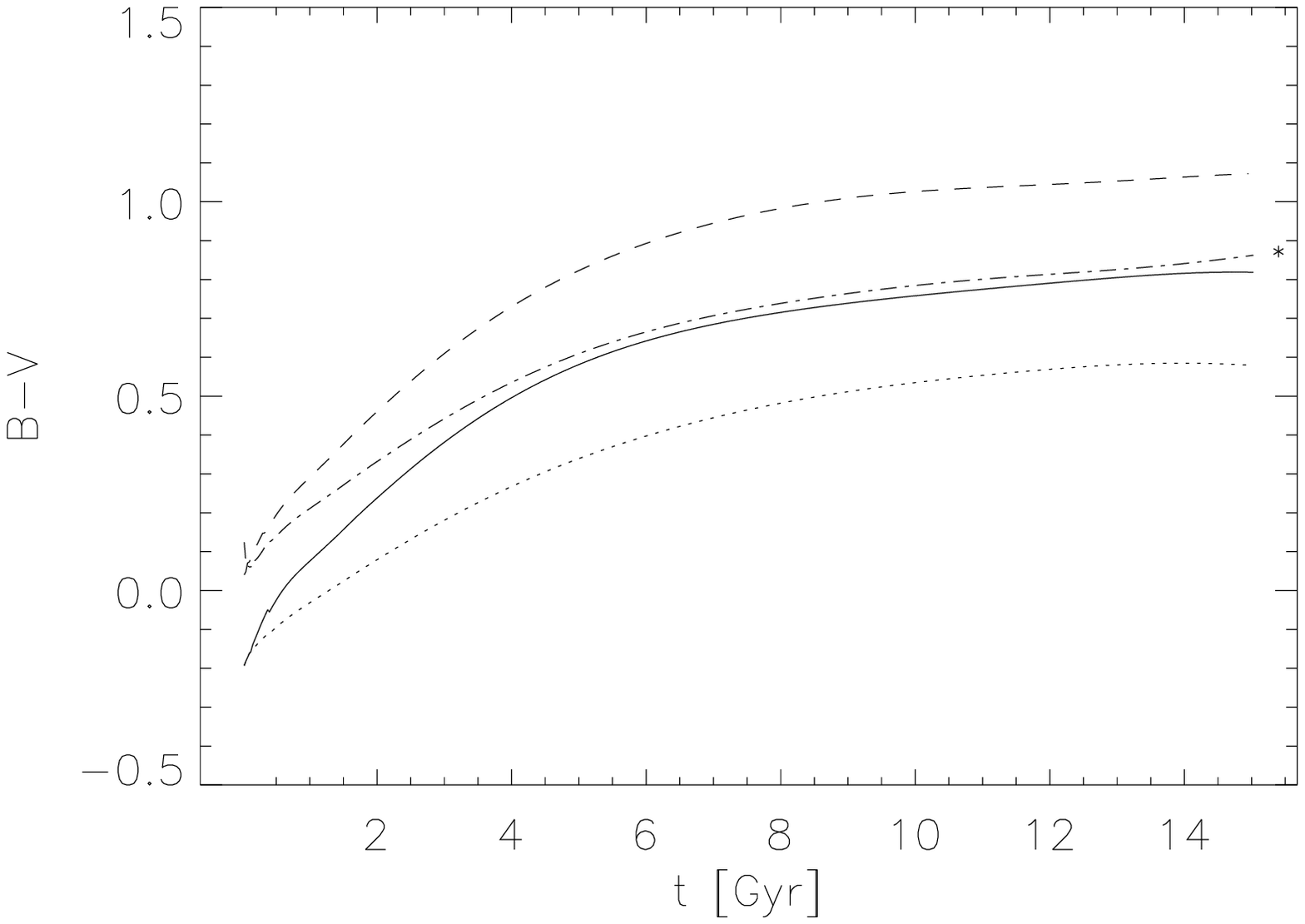}}
\epsfysize=6 cm
{\epsffile{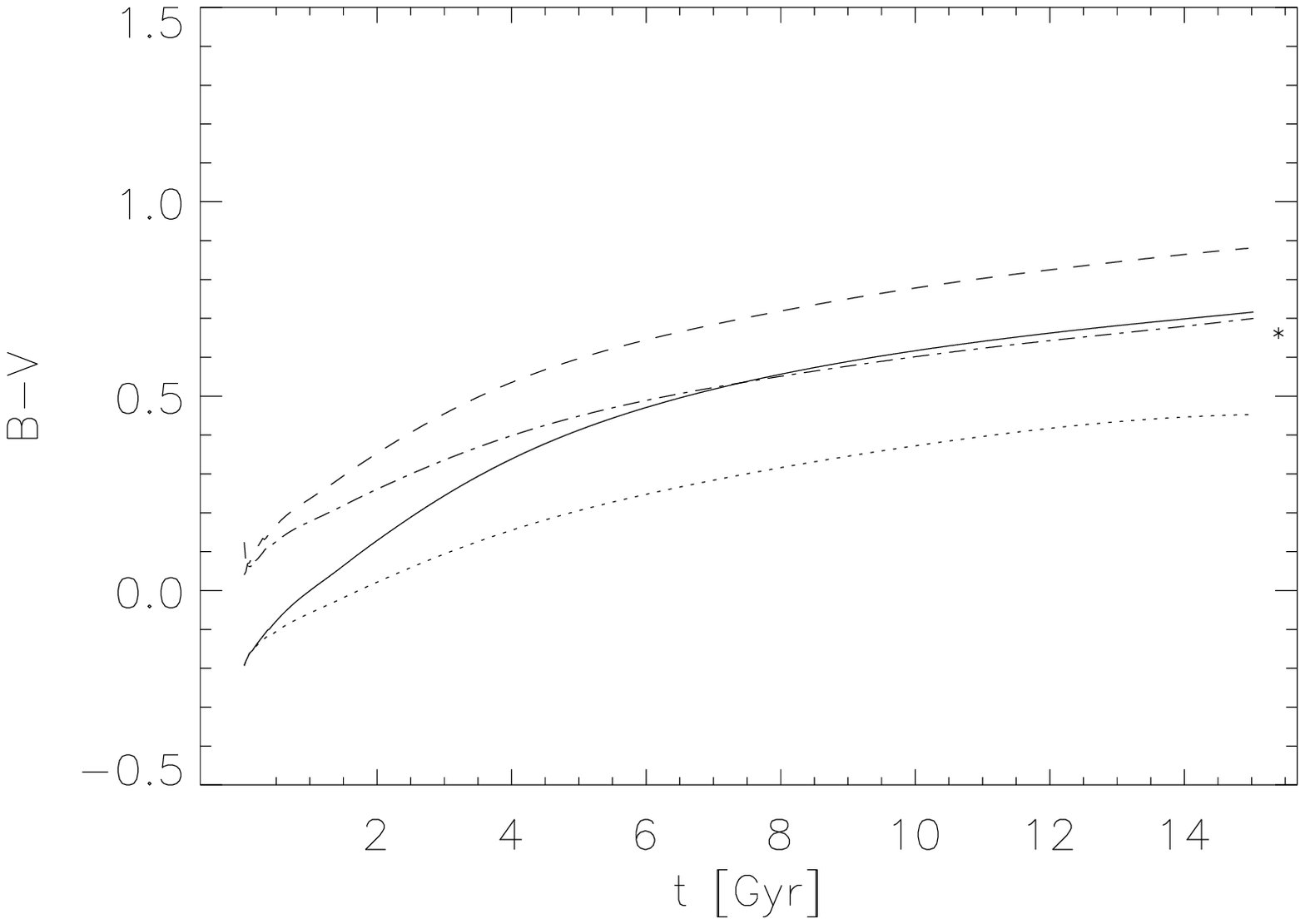}}
\epsfysize=6 cm
{\epsffile{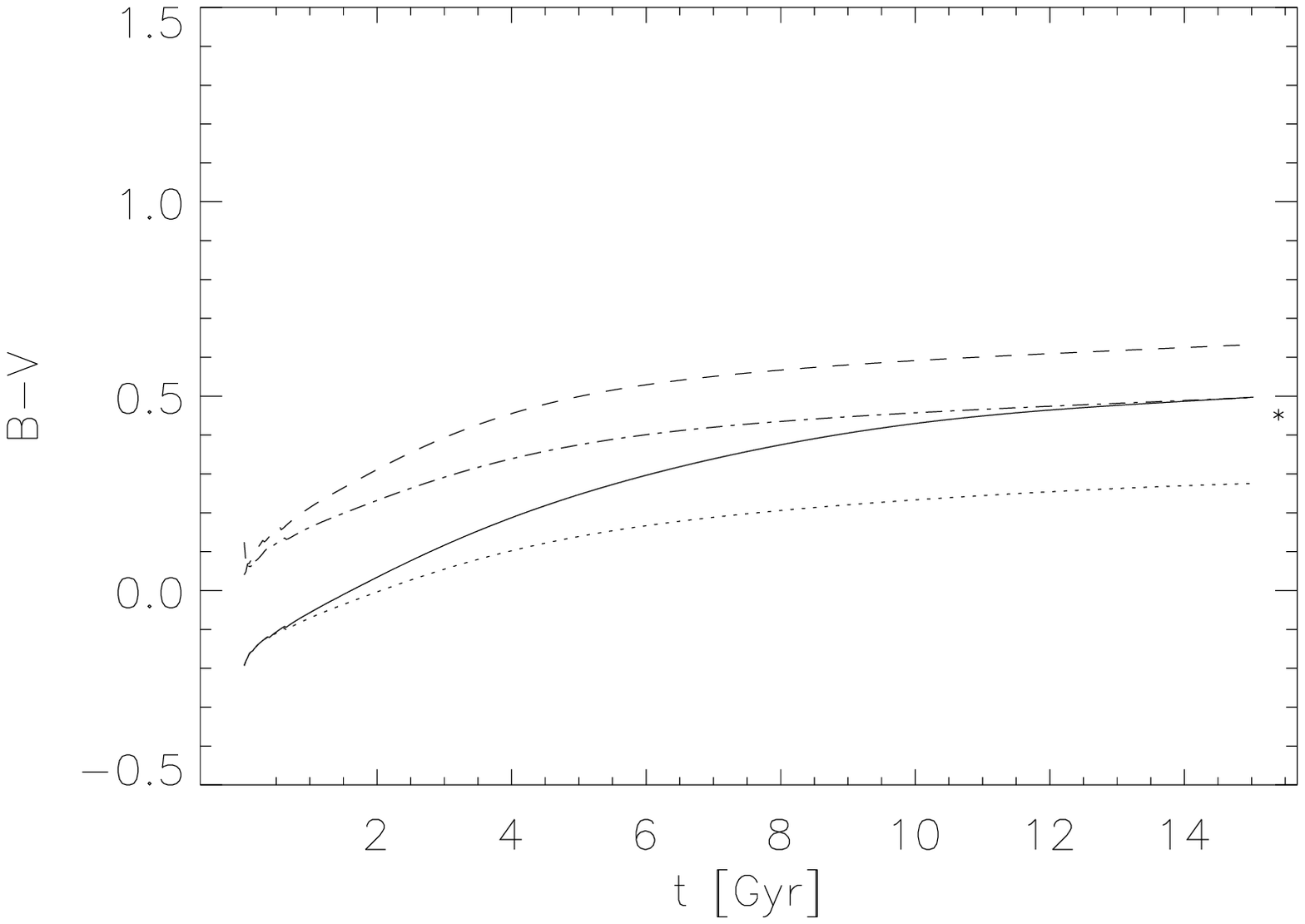}}
\figure{1}{Color evolution (B-V) of different spectral types: elliptical 
(top panel), Sb (middle panel) and Sd (bottom panel). 
Lines are solid for the chemically consistent model, 
dotted for $Z= 10^{-4}$, dash - dotted for $Z=4 \cdot 10^{-3}$ and dashed for
$Z= 2 Z\odot$. The data points (*) are the observations from de Vaucouleurs
et al.  (1991 RC3).}
\endfig

\begfig 18 cm
\vskip -18 cm
\epsfysize=6 cm
{\epsffile{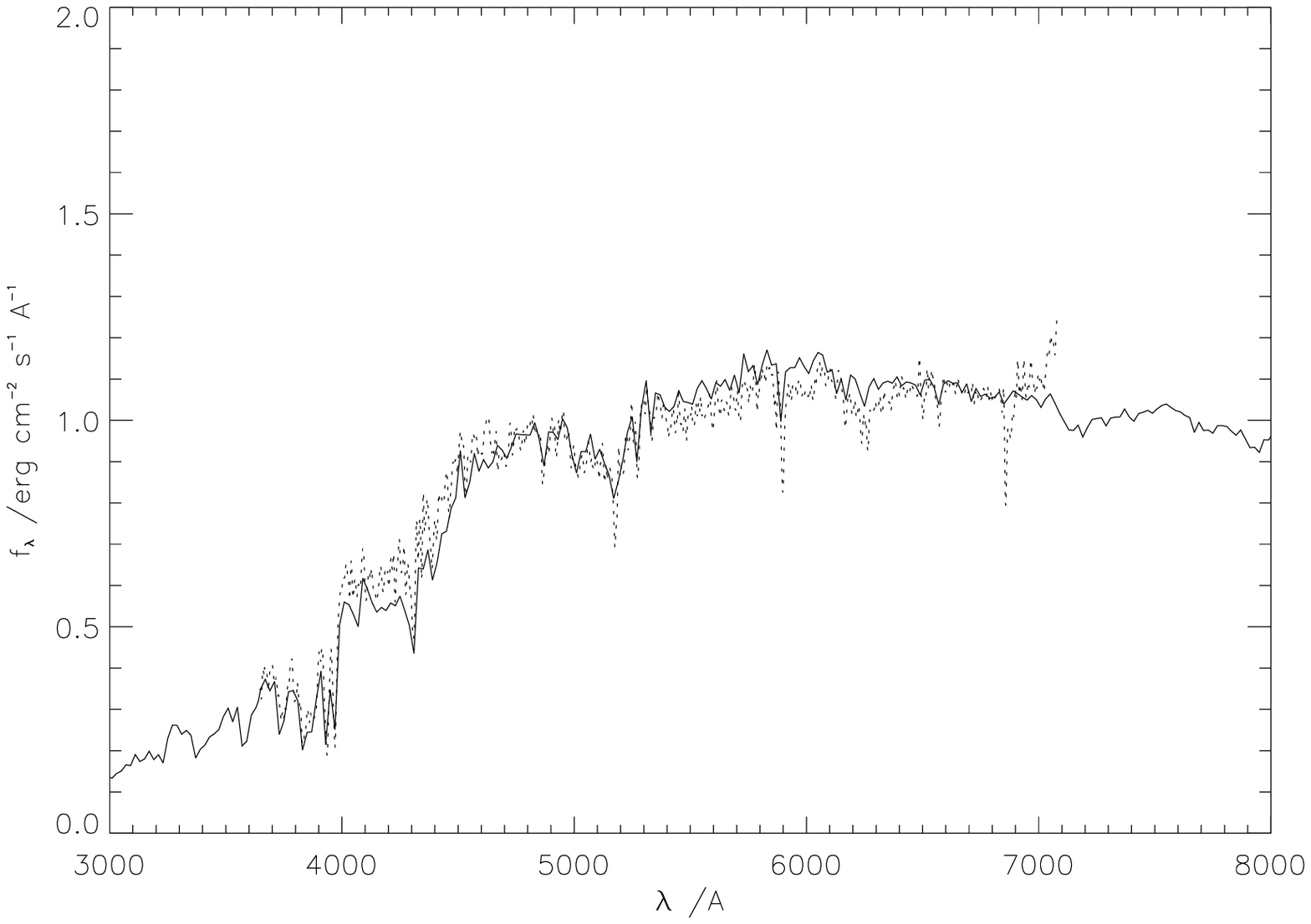}}
\epsfysize=6 cm
{\epsffile{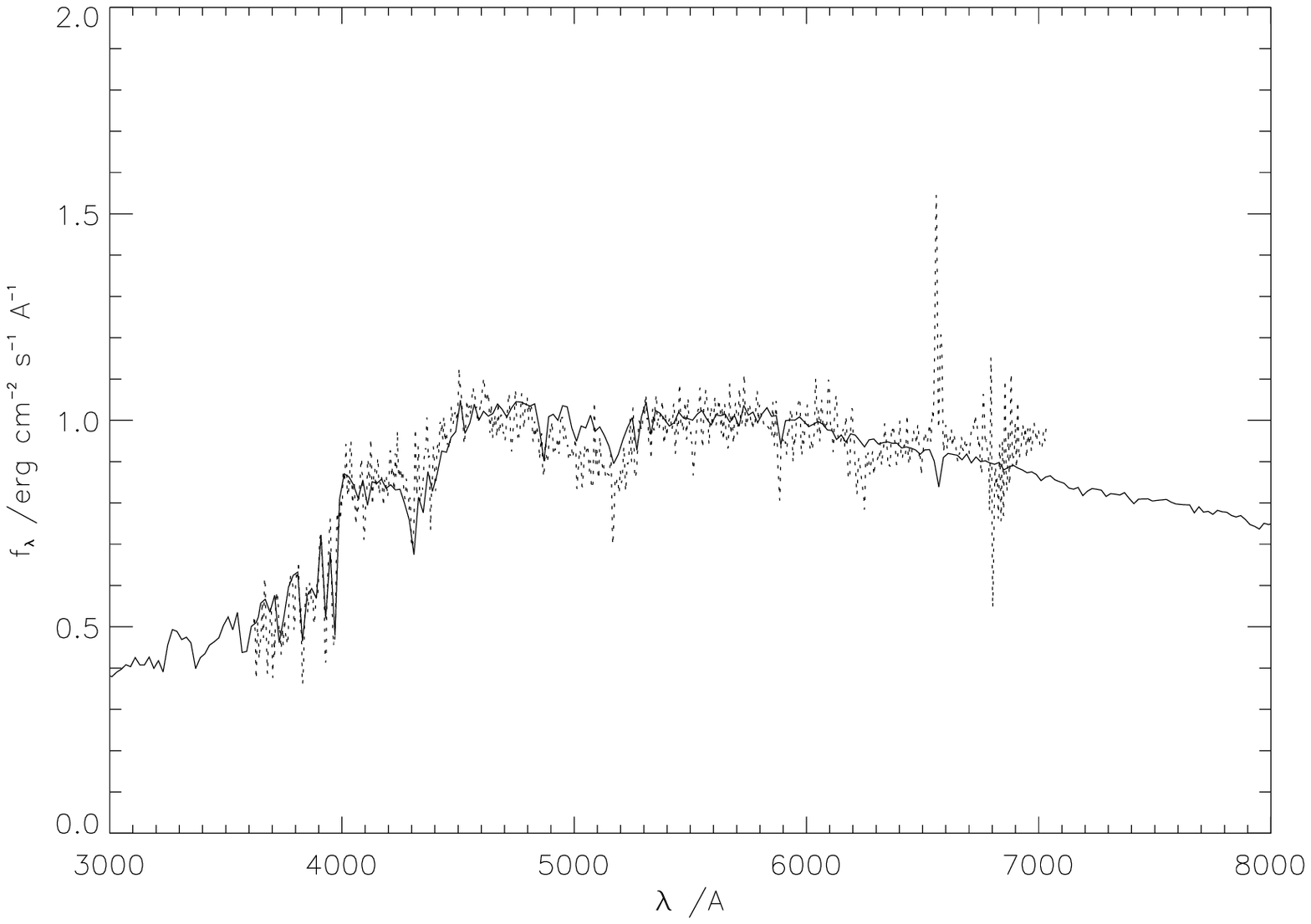}}
\epsfysize=6 cm
{\epsffile{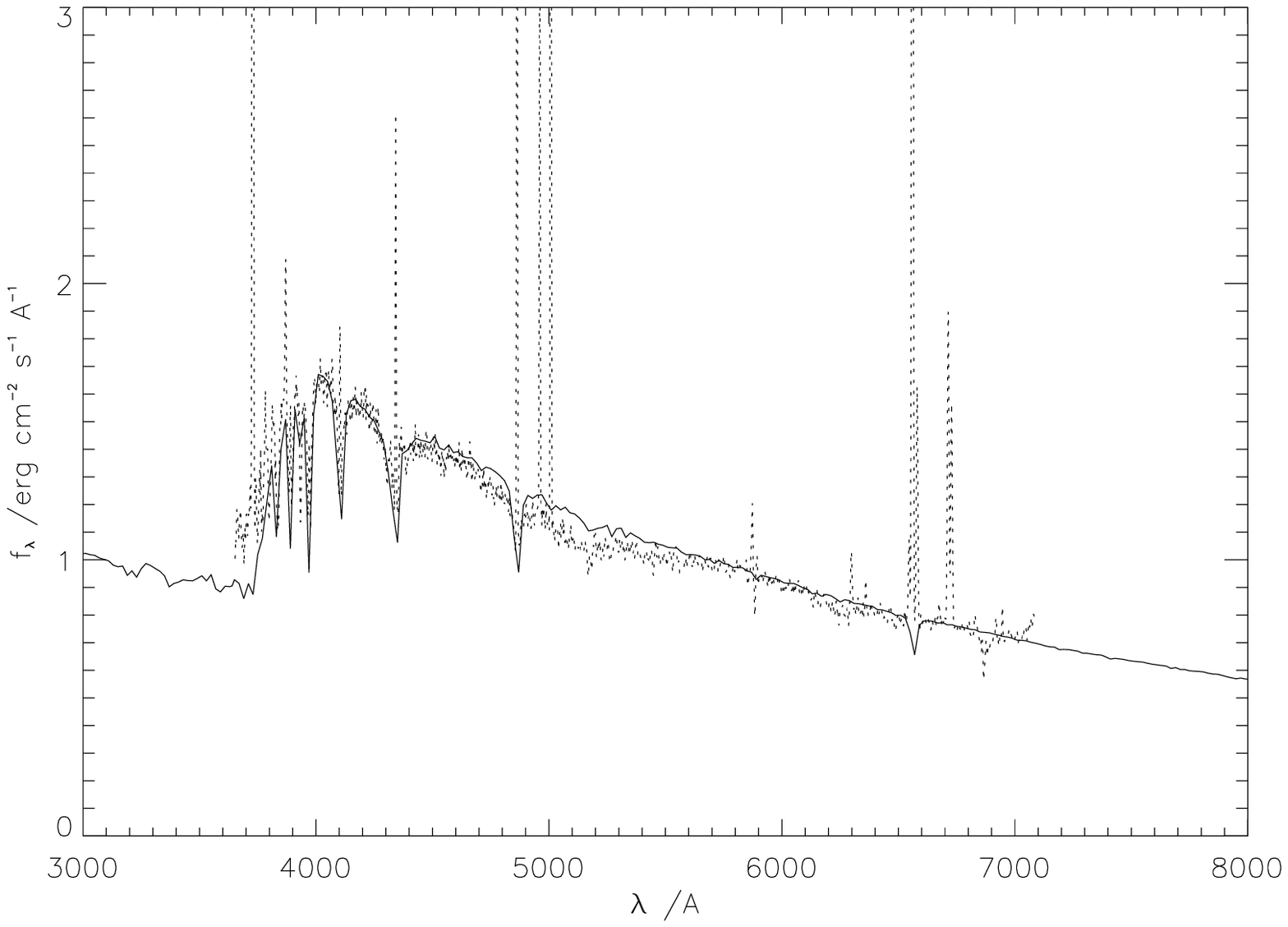}}
\figure{2}{Comparison of theoretical galaxy spectra (solid) with 
observations from Kennicutt (1992) (dotted). NGC 3379 (E0) with E - model 
of 11 Gyr (top panel), NGC 3147 with Sb - model of 12 Gyr (middle panel)
and NGC 4449 with Sd - model of ~ 8 Gyr (bottom panel).}
\endfig

To visualize the metallicity effects on the 
color evolution, we plot in Fig. 1 the  
chemically consistent model, together with three single metallicity
models ($Z= 10^{-4}, 4\cdot 10^{-3}, 4\cdot 10^{-2}$) and their evolution in 
time.
To justify the use of our SF laws (Tab. 1) together with our Scalo IMF for 
the different spectral types, we show that the chemically consistent
calculations reproduce the
color observations from de Vaucouleurs et al. (1991 RC3) of
typical ellipticals (top), Sb - spirals (middle) and Sd - galaxies 
(bottom panel). As expected, the color
evolution at constant metallicity for all Hubble types
is bluer at lower metallicities.
It is clear that models with tracks of solar
metallicity are redder than our chemically consistent model and that this 
difference is even more pronounced at earlier times. In our E - model 
the color evolution
starts, of course, from the lowest metallicity, reaching the curve
for $Z=4\cdot 10^{-3}$ after $\sim $4 Gyr and keeping
this color until 15 Gyr. The 
mean stellar metallicity stays subsolar throughout a Hubble time. In Sb-type
galaxies after 8 Gyr, the (B-V) 
curve for the chemically consistent model approaches 
the one that only uses
evolutionary tracks with $Z=4\cdot 10^{-3}$.
Because of their constant star formation rate Sd - galaxies evolve very 
slowly towards redder colors. 
The chemically consistent (B-V) color of Sd's only reaches 
that of a single metallicity population with $Z=4\cdot 10^{-3}$
after 11 Gyr.  

In calculations with a single constant metallicity, stellar tracks with
$Z=4\cdot 10^{-3}$ seem to produce colors in better agreement with 
observations than models with solar 
metallicity. After a Hubble time 
the chemically consistent method gives (B-V) values,
depending on the star formation rate, between
0.2 and 0.5 mag bluer than that of the
model with solar metallicity. 

For another test of our models we compared our integrated galaxy spectra 
with observations from a sample 
of 90 nearby ($z \le 0.03$) galaxies of various spectral types from Kennicutt (1992). 
The observed galaxy spectra cover a wavelength range 
from 3650 - 7000 \AA and the spectral
resolution varies from 5 \AA \ to 25 \AA, while our model spectra have a 
resolution of 20 \AA \ in this range. 
 
In Fig. 2 we 
compare some of Kennicutt's template spectra 
with our synthetic model spectra of the respective Hubble types.
The model spectra are computed without gaseous emission lines.
Fig. 2a) shows NGC 3379 (=E0 template) with our elliptical model, b)
NGC 3147 (Sb) with our Sb-model and c) NGC 4449 an irregular galaxy (Sm/Im)
which seems to be well described by the constant SFR of our Sd-model. The 
spectrum of NGC 3379 is fitted well by our 11 Gyr old E - model. Because the spectral
energy distribution of ellipticals evolves strongly during the first 12 Gyr, 
the age of
NGC 3379 can be restricted to a small range. The spectrum of a young elliptical
declines at $\lambda \ge 6000$ \AA, while the Balmer jump 
becomes higher in older E - galaxies because 
of the accumulation of low mass main - sequence 
stars and cold giants.
NGC 3147 is well reproduced by a 12 Gyr old Sb - spectrum. Our Sd - models
with an age of 8 -9 Gyr are suitable for the NGC 4449 spectrum. Because of a 
constant
star formation rate the spectrum of an Sd - galaxy doesn't evolve very 
much, resulting in an
age dating, that is less precise than for early type galaxies.
The time dependent composition of the five stellar tracks produces the smooth
evolution of the Sd spectrum.

Bruzual \& Charlot (1993) and Bruzual (1993)
produced solar metallicity models and fit the
observed spectrum of NGC 3379 with a 13.8 Gyr old elliptical spectrum,
that of NGC 3147 with an 8 Gyr old Sb - spectrum and that of NGC 4449
with a 1 Gyr young Sd - spectrum (constant star formation rate).
This shows that the strong subsolar metallicity of late type galaxies 
leads to 
erroneously low ages when determined from a solar metallicity model.

For a more detailed description of our models see M\"oller (1995),
where all colors are calculated from the U to K band, and 
the color evolution over a large redshift range, assuming a set of cosmological
parameters, is compared to observations of high redshift galaxies.

\titlea{Metallicity indicators in different wavelength regions}

\titleb{Evolution of ISM abundances}

In Fig. 3 we show the enrichment of the ISM metallicity Z for the star
formation histories ascribed to the different galaxy types.
The
horizontal lines mark the five metallicity regimes
and delineate the critical ISM metallicities at which we change, 
for the newly born stars, from one set of stellar
evolutionary tracks to another.

\begfig 6 cm
\vskip -6cm
\epsfysize=6cm
{\epsffile{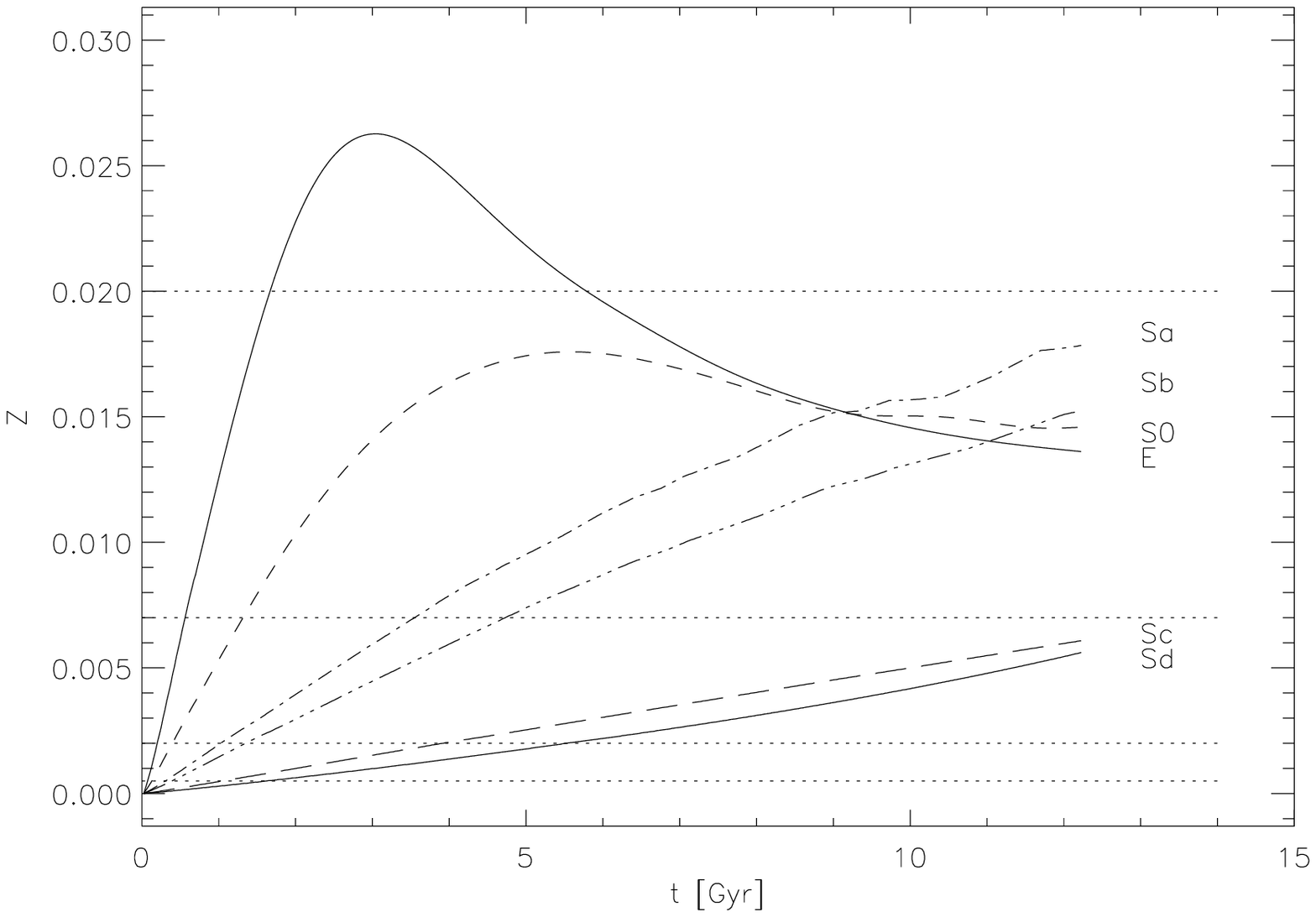}}
\figure{3}{Time evolution of ISM metallicity for various spectral types. 
The horizontal lines mark the five discrete metallicity regimes.}
\endfig

The metallicity of 
spiral galaxies increases steadily with time. The enrichment is
faster and stronger for earlier spiral types because of their higher
star formation rates. Owing to the very strong star formation in the first Gyr,
the ISM abundance in ellipticals increases strongly up to ~3 Gyr. At this time,
virtually all of the gas is consumed by star formation and most of the massive stars
have already died out. After that the lower mass stars begin dying in large
numbers, increasing the total gas content but mainly giving back unenriched material.
This accounts for the well-known dilution effect of the 
ISM in ellipticals. The strength of the dilution effect depends on the ratio
of low mass stars to high mass stars given as by the IMF. While a dilution effect is 
clearly seen for a Scalo IMF, a Salpeter IMF with its lower ratio of low - to - high
mass stars only causes a leveling - off of the ISM metallicity for $t \ge 3$ Gyr.
Our model ISM abundances agree with observations (e.g. Zaritsky et al. 1994) 
for the various galaxy types.
Because of their significantly subsolar metallicity, the effects of chemical
enrichment will be more important and visible in the
spectrophotometric appearance of late type galaxies. 
Except for the central regions of massive ellipticals, the 
average stellar metallicities
in dwarf and normal ellipticals are expected to be subsolar as well.

\titleb{Luminosity weighted stellar metallicities}

As described above, 
a major problem with composite stellar populations is  
that the same broadband color can be produced by young metal - poor 
stars as well as by
old metal - rich stars (e.g. O'Connell 1976, Arimoto \& Yoshii 1987, 
Bica et al. 1988, Faber et al. 1995).
It is likewise well known that the NIR is much less age dependent than the optical 
or the UV.
For this and other reasons metal indices are investigated in the NIR 
(e.g. Frogel 1985,
Terndrup et al. 1991, Lan\c con \& Rocca - Volmerange 1992). 
We analyse the luminosity weighted stellar metallicities
as a function of wavelength because their 
behaviour is important for the comparison of metal indices in different bands.
Metallicities deduced from observations are always luminosity weighted,
in contrast to mass-averaged metallicities of composite stellar populations
(Matteucci 1994). 

We use the following definition for our calculations of the luminosity weighted
stellar metallicities:
$$ Z_{\lambda}= a_1 \cdot Z_1+ a_2 \cdot Z_2+ a_3 \cdot Z_3+  a_4 \cdot Z_4+ 
 a_5 \cdot Z_5      \eqno(1) $$
where $a_i := {L_{\lambda}(Z_i)  \over L_{\lambda tot}} $ is the relative 
contribution to the luminosity $L_{\lambda}$ in the band $\lambda $ from stars
with metallicity $Z_i$.

This luminosity weighted stellar metallicity can differ greatly from a 
mass-weighted one. Arimoto \& Yoshii (1987) and Matteucci \& Tornamb\`e (1987)
have already shown that, in general, the luminosity averaged metallicity in visual
light is smaller than the mass-weighted one, since metal poor giants dominate 
this wavelength region.
It is worth noting that in simple stellar populations
$Z_{\lambda}$ is the same in all wavelengths 
and that differences in $Z_{\lambda}$ can only be described by
a chemically consistent evolutionary synthesis.

The evolution of the gas metallicity reflects the star formation history,
but is not
linearly correlated with the metallicity of the stellar population. 
ISM abundances are normally obtained from HII regions.

\begfig 6 cm
\vskip -6.5cm
\epsfysize=6cm
{\epsffile{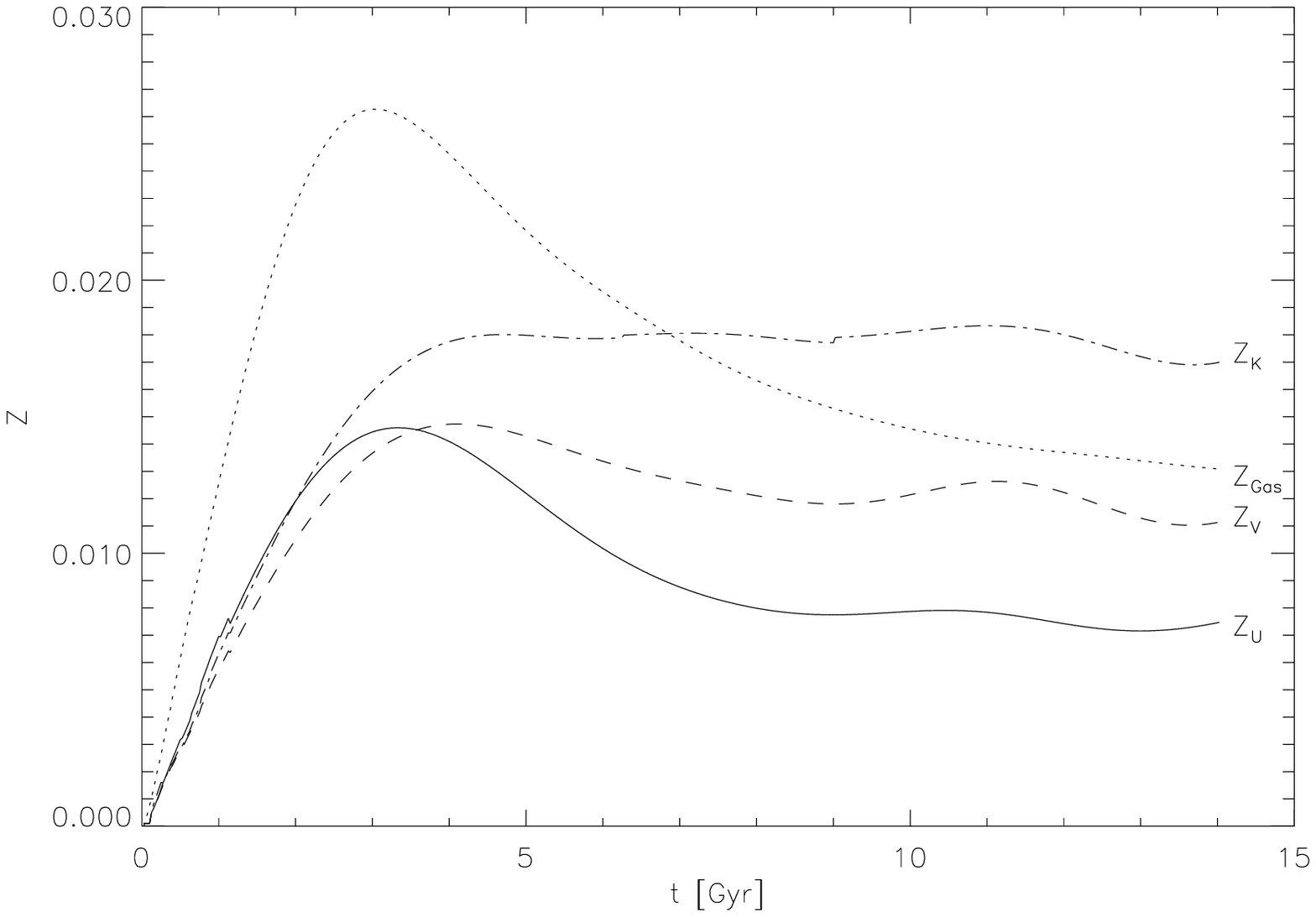}}
\figure{4}{Time evolution of luminosity weighted stellar metallicities $Z_U$(solid),
$Z_V$(dashed) and $Z_K$(dash-dotted) and of ISM metallicity (dotted) for ellipticals.}
\endfig

\begfig 6 cm
\vskip -6.5cm
\epsfysize=6cm
{\epsffile{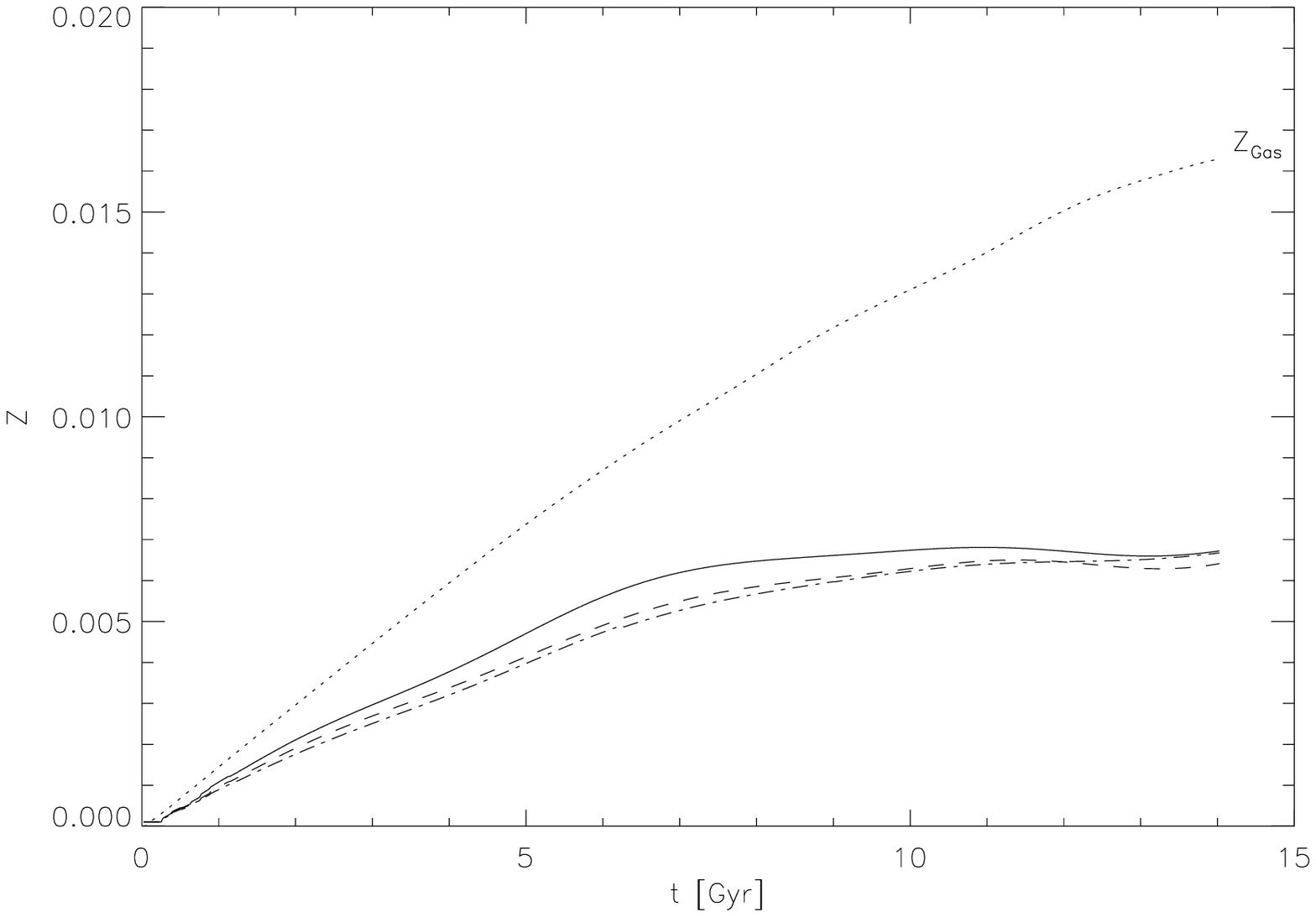}}
\figure{5}{Same as in Fig. 4, but for Sb - galaxies.}
\endfig

\begfig 6 cm
\vskip -6.5cm
\epsfysize=6cm
{\epsffile{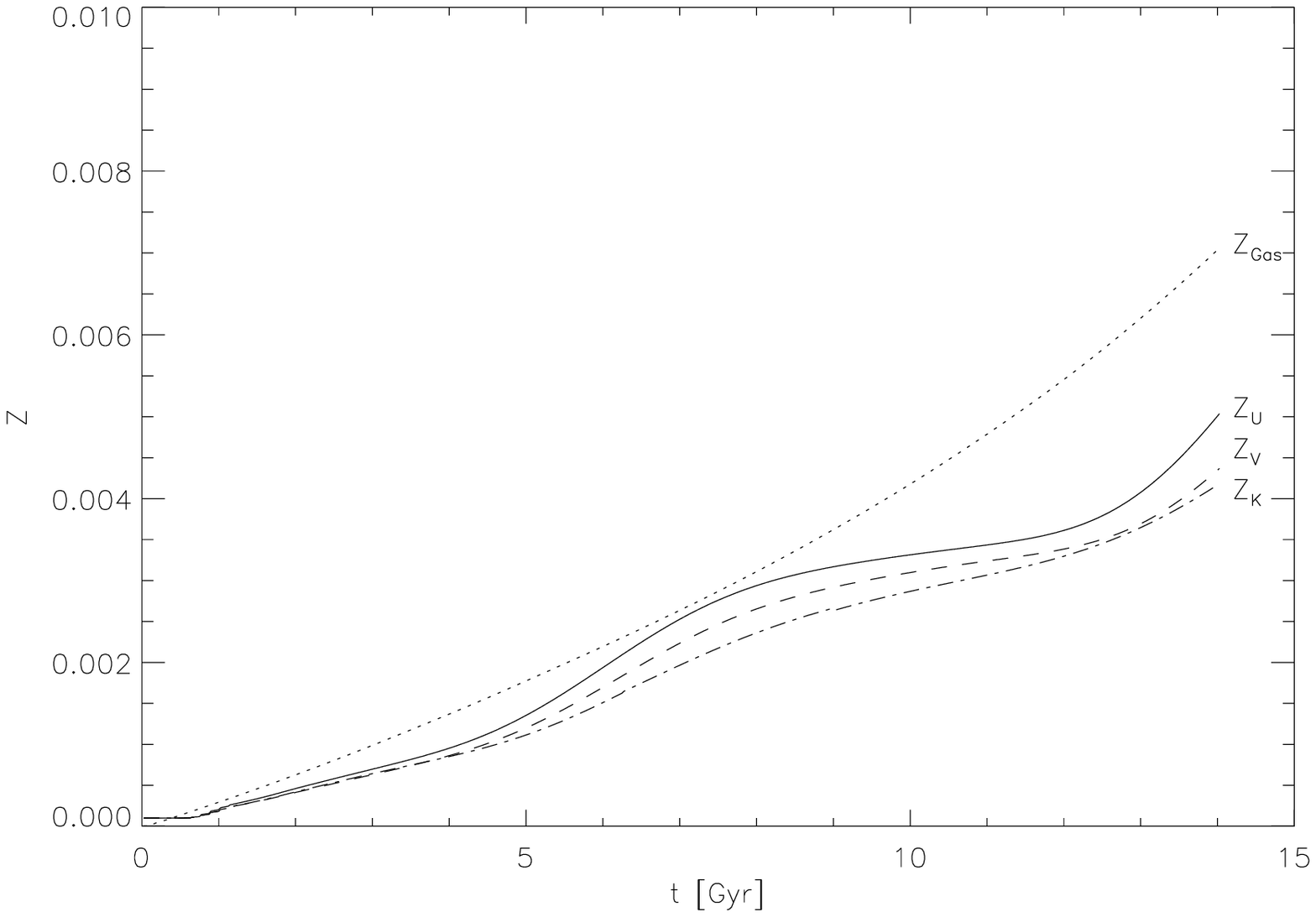}}
\figure{6}{Same as in Fig. 4, but for Sd - galaxies.}
\endfig

\begfig 6 cm
\vskip -6.5cm
\epsfysize=6cm
{\epsffile{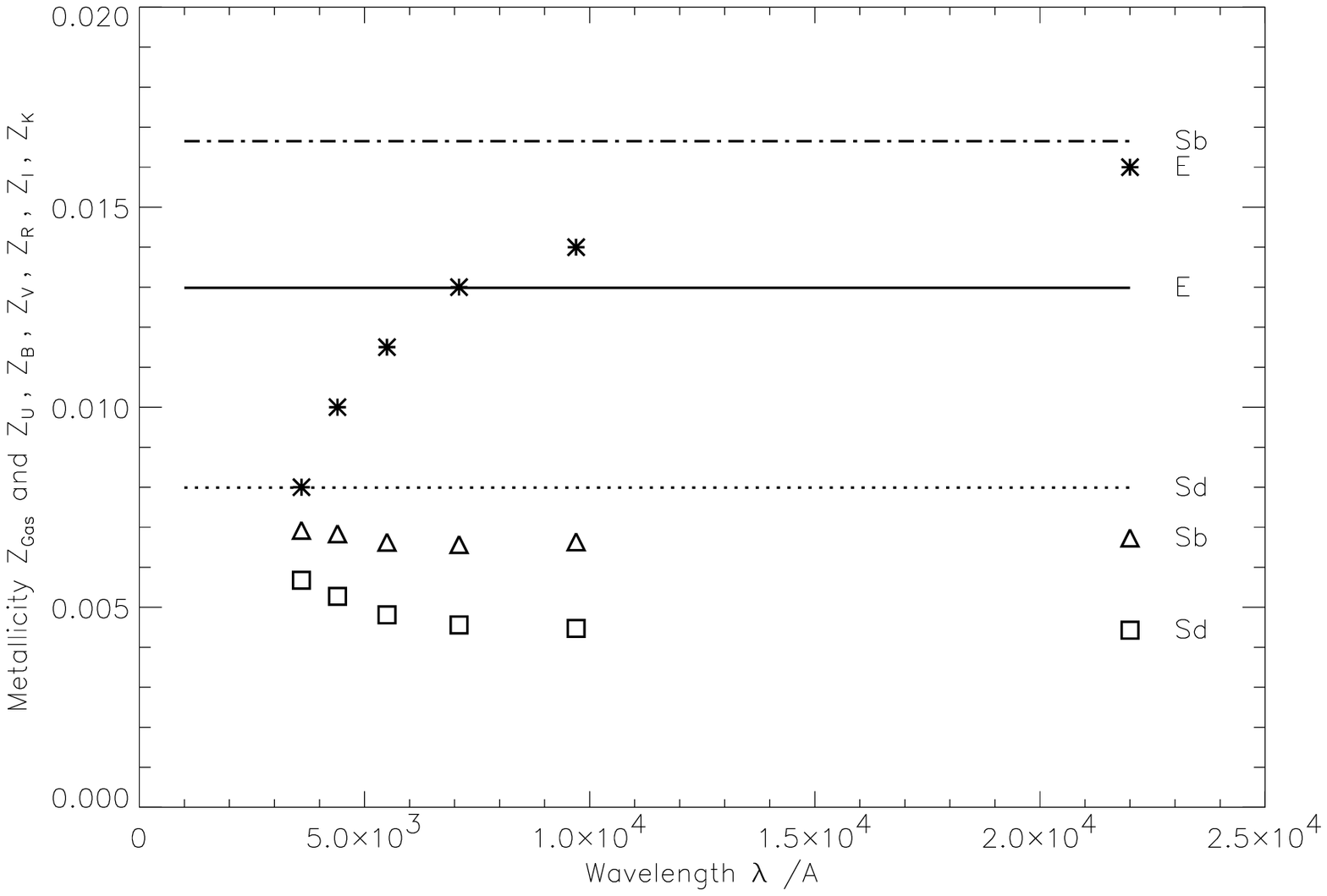}}
\figure{7}{Comparison of stellar metallicities $Z_U$, $Z_B$, $Z_V$, $Z_R$, $Z_I$,
 $Z_K$ with gas metallicity (horizontal lines) after 15 Gyr for E (star), 
 Sb (triangle) and Sd (box).}
\endfig

In Fig. 4 - 6 we compare the time evolution of ISM metallicities to the luminosity
weighted stellar metallicities for galaxy 
types E, Sb and Sd. In ellipticals (Fig. 4) the
enrichment of the metallicity is faster in the gas than in the 
stellar population. Like the ISM abundance, all stellar
metallicities $Z_U ... Z_K$ increase during the first 3 - 4 Gyr. 
Only $Z_U$, which is dominated by massive young stars, decreases after a 
maximum, in a very similar manner to the dilution in the gas. 
Any stars formed at times later than $\sim $4 Gyr will again have lower initial
metallicities. Very different is the behaviour of $Z_K$ which nearly stays
at its maximum. $Z_K$ mainly represents the
average metallicity of cold, low - mass main sequence stars and red giants which
give strong contributions to the luminosity in K. After 7 Gyr 
the gas metallicity decreases
below the value of the stellar metallicity in K because of
the above mentioned dilution effect. The mean mass - weighted stellar 
metallicity is lower
than $Z_K$ and obviously cannot exceed the gas abundance.

The stellar metallicities in Sb - spirals (Fig. 5) steadily
increase with time. There is very little difference between luminosity 
weighted stellar metallicities at various wavelengths from optical to UV. 
$Z_U$ through $Z_K$ stay below the ISM metallicity by a factor  
of two throughout a Hubble
time. For Sd type galaxies, with their star formation rate being constant in time,
stellar metallicities from U through K are very similar and only differ slightly
from the global ISM metallicities (Fig. 6). 

Figs. 4,5 and 6 clearly show that stellar metallicity indicators at different
wavelengths and HII - regions abundances have to be
expected to give significantly different results. 
Our models quantify these differences and explicitly show how these depend on the
SF history.

To analyse the characteristic differences, we show
in Fig. 7 the luminosity weighted stellar metallicities 
$ Z_U, Z_B, Z_V, Z_R, Z_I, Z_K$ and the
ISM abundance for different galaxy types after 15 Gyr. Going from the U to
the K band the trend for luminosity weighted stellar metallicities in ellipticals
is just the inverse to that for spirals. Even within a single galaxy, 
metal abundances derived
from different wavelength regions 
are predicted to differ by as much as a factor of two. 
The metal enrichment of the stars differs from that of the gas 
by a factor of two in Sb - galaxies, more
than in any other type of galaxy (M\"oller et al. 1995).
This explains the differences between both the metallicities and Galactic metallicity
gradients derived from HII - region 
abundance studies and metallicities individually determined for samples of 
B-, G-, K- or M stars which emit their dominant light contributions at different
wavelengths (Nissen 1995, Kilian et al. 1994, Edwardsson et al. 1993).  

\begtabfull
\tabcap{2}{Percentage luminosity contribution to 
     the total luminosity $L_{tot}$[$10^9 L_{\odot}$] in U, V, K bands 
     from stellar subpopulations of different metallicities at 15 Gyr.}
\halign{#\hfil&\quad\hfil#\hfil&\quad\hfil#\hfil&\quad\hfil#
         \hfil&\quad\hfil#\hfil&\quad\hfil#\hfil&\quad\hfil#\hfil\cr
\noalign{\medskip}
 Z =  & $10^{-4}$ & $10^{-3}$ & $4\cdot 10^{-3}$ & $10^{-2}$ & $4\cdot 10^{-2}$ & $L_{tot}$ \cr
\noalign{\medskip\hrule\medskip}
      &           &           &  E          &           &             &           \cr
\noalign{\medskip\hrule\medskip}
$L_U$ &    14.9    &    16.8  &   38.5      &    28.6   &      1.2    &    16.1   \cr
$L_V$ &    6.5     &    10.2  &   29.0      &    52.2   &      2.1    &    24.5   \cr
$L_K$ &    2.7     &    5.1   &   14.7      &    75.6   &      1.9    &    81.9   \cr
\noalign{\medskip\hrule\medskip}
      &           &           &  Sb         &           &             &           \cr
\noalign{\medskip\hrule\medskip}
$L_U$ &    5.1    &    5.6    &     22.3    &    67.0   &       -     &    39.0   \cr
$L_V$ &    3.0    &    3.9    &     30.0    &    63.1   &       -     &    43.6   \cr
$L_K$ &    1.6    &    2.8    &     31.4    &    64.3   &       -     &   116.1   \cr
\noalign{\medskip\hrule\medskip}
      &           &           &  Sd         &           &             &           \cr
\noalign{\medskip\hrule\medskip}
$L_U$ &    0.9    &   18.7    &    80.4     &      -    &       -     &   158.0   \cr
$L_V$ &    1.1    &   15.5    &    83.4     &      -    &       -     &    91.3   \cr
$L_K$ &    1.2    &   10.9    &    87.9     &      -    &       -     &   120.6   \cr
\noalign{\medskip\hrule}}
\endtab
    
To further 
explain these results, we show in Table 2 the percentage luminosity 
contributions to the U, V, K filters from stellar subpopulations of different 
metallicities at an age of 15 Gyr. Column 7 contains, respectively, the total
luminosity $L_U$, $L_V$, $L_K$ in [$10^9 L_{\odot}$].
The dominant contribution to the K luminosity in 
ellipticals is from  stars with metallicities of about 0.5 Z$_\odot$. 
This stellar subpopulation which accounts for
$2\!/3$ of the total light in K, represents at the same time, the bulk of the 
stellar  mass in normal ellipticals. The luminosity distribution in V is somewhat
broader than in K, while the broadest metallicity 
distribution is found in U.   
The luminosity in U is mainly due to late forming massive stars with subsolar
metallicities. The maximum of the metallicity distribution depends on wavelength, 
approximately $Z=4\cdot 10^{-3}$ in U and up to $Z=4\cdot 10^{-2}$ in the K band.
Stars with $Z \sim 2 Z_{\odot}$ do not significantly contribute to the 
integrated U, V and K light.
While the metallicity distribution of stars contributing light to some band
reaches a maximum and then 
declines towards the highest metallicities in each band
for ellipticals, it shows a monotonic, though overall weaker, increase in
spirals.
There are  
characteristic differences between early and late type spirals. In Sb - galaxies, 
$L_K \ge L_U, L_V$ for every metallicity subpopulation except 10$^{-4}$, while in
Sds, $L_U \ge L_V, L_K$ for all metallicities.
As expected from Fig. 1, the maximum metallicity reached by stars is 0.5 $Z_{\odot}$
in an Sb - spiral and $Z=4\cdot 10^{-3}$ in an Sd - galaxy.


\titleb{Discussion}

In 1988, Bica et al. already presented an analysis of the V light contributions
of different metallicity components in nuclei and entire galaxies of various star
formation histories. To do this, they combined star cluster population synthesis with
M/L$_V$(t) from Arimoto \& Yoshii's evolutionary synthesis. Using a very flat 
IMF (x= 0.95), their model
results in a stronger enrichment and consequently gives stronger 
V light contributions of high metallicity components as compared to our model.

We want to recall that, though it's not possible in our model to constrain the IMF
and the SF histories independently of each other, our combination of a uniform IMF 
with empirical SF histories as given in Table 1, yields ISM abundances that agree 
with those observed by Zaritsky et al. (1994) in a large sample of galaxies.

To conclude, we find that the dominant light contributions to the various bands
in global galaxy models are from stars of $\sim 0.5 Z_{\odot}$ metallicity for
normal gEs to Sb galaxies, while in Sd galaxies it comes from stars of lower 
metallicity. The situation may of course be different in the nuclei or bulges of
galaxies, in dwarf ellipticals and cD galaxies, for which our simple one-zone, 
closed box models do not apply.

\titlea{Metal Indices}

\titleb{Time evolution and relations with [Fe/H]$_*$}

\begfig 6 cm
\vskip -6.5cm
\epsfysize=6cm
{\epsffile{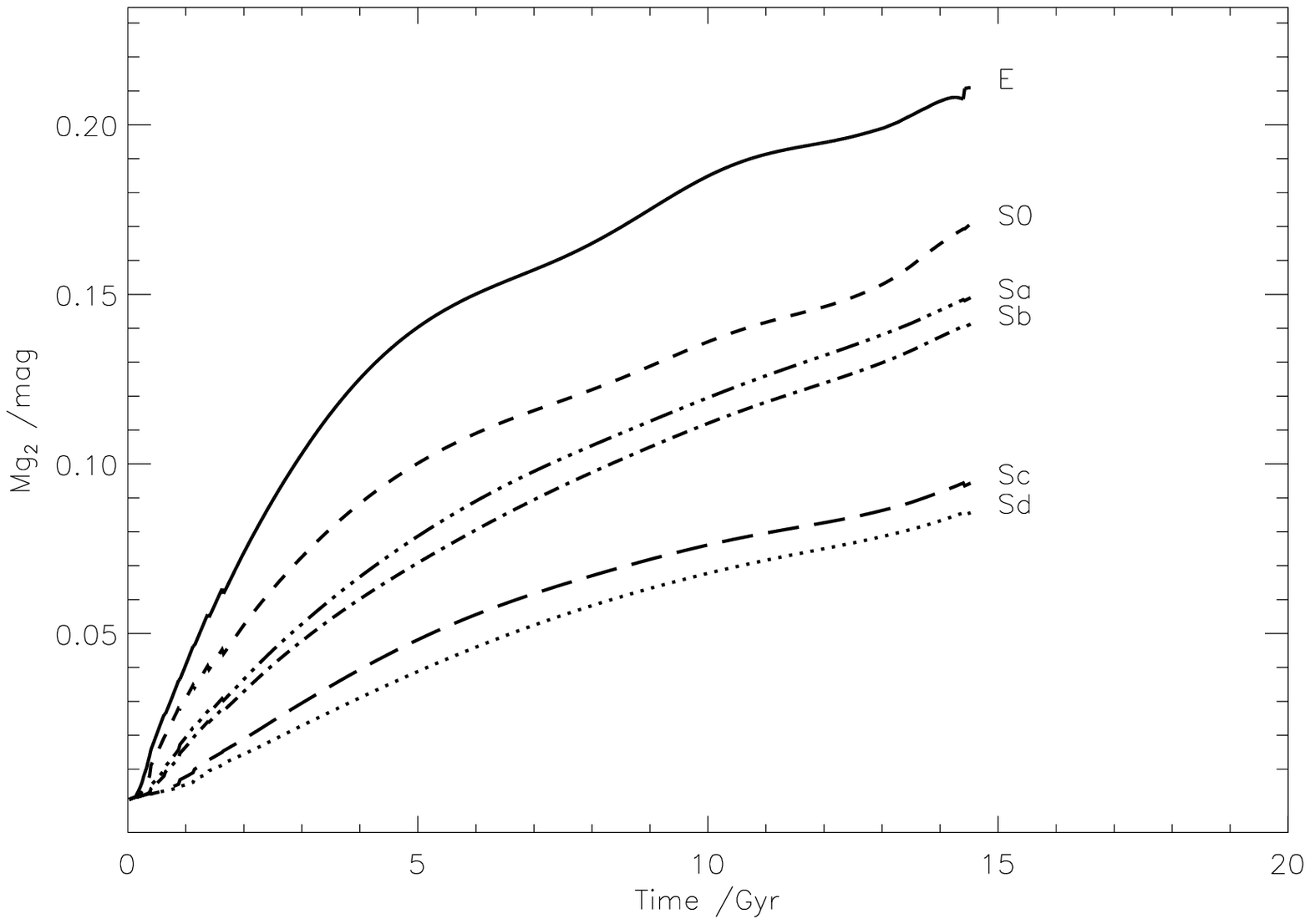}}
\figure{8}{Time evolution of $Mg_2$ for galaxy models of different spectral types.}
\endfig

\begfig 6 cm
\vskip -6.5cm
\epsfysize=6cm
{\epsffile{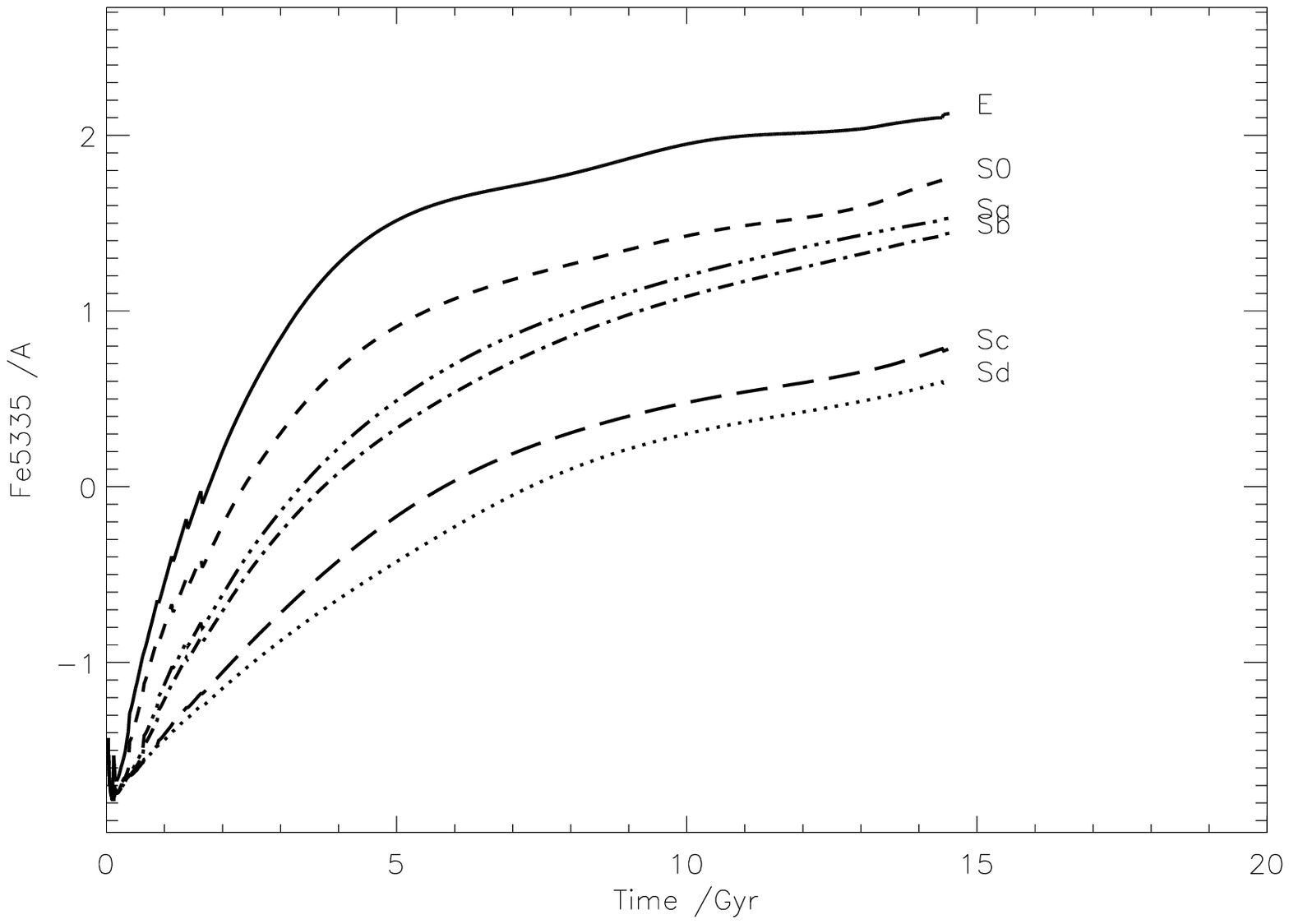}}
\figure{9}{Time evolution of Fe5335 for galaxy models of different spectral types.}
\endfig

\begfig 6 cm
\vskip -6.5cm
\epsfysize=6cm
{\epsffile{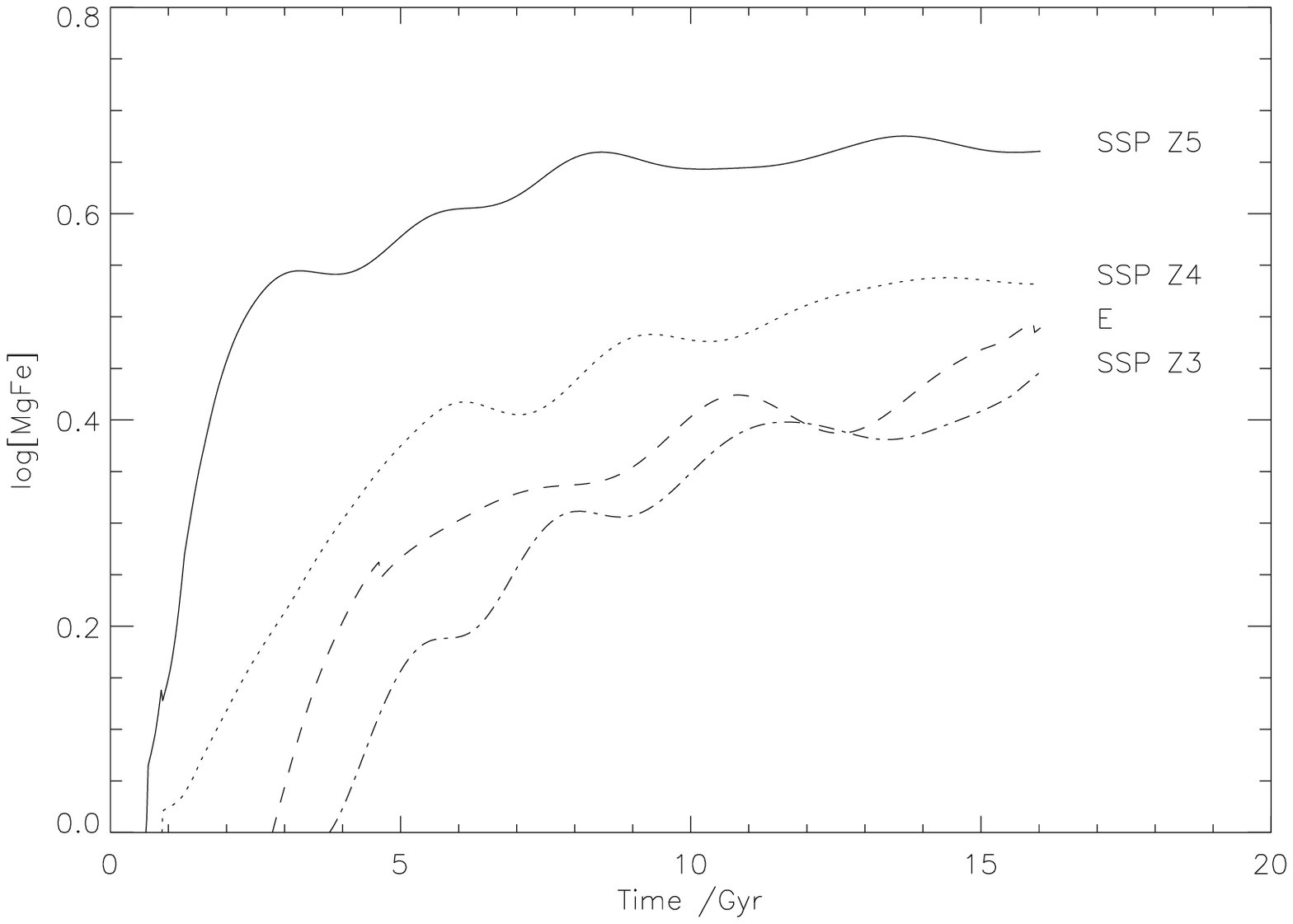}}
\figure{10}{Time evolution of stellar index [MgFe] for an E-galaxy model (dashed)
and for three SSPs with Z3=$4\cdot 10^{-3}$ (dot-dashed), 
Z4=$10^{-2}$ (dotted) and Z5=$4\cdot 10^{-2}$ (solid). The range of observations 
(Gonzales 1993) is given by the perpendicular line.}
\endfig

Integrated stellar metallicities of galaxies are generally determined from 
observations of characteristic stellar absorption features (e.g. Mg$_2$, Mgb,
iron lines, TiO bands, etc.). Unless reliable emission line subtraction can be
performed, these metal indices can only be measured from emission free spectra of 
early type galaxies. As explained in Sec. 2.2, we calculate the time evolution of
a series of stellar metallicity indicators for our model galaxies. For the case of 
early type galaxies these can be directly compared to observations.

We show the time evolution of $Mg_2$ and Fe5335 in
Figs 8 and 9 for the Hubble sequence of galaxies from E to Sd. 
As expected, 
the stellar magnesium and iron indices increase 
continuously for all galaxy types. These indices
rise faster and reach higher values 
in ellipticals, although their final gas metallicity is lower than in early type spirals
after a Hubble time.
The metallicity indices we analyse represent the metallicity of the stellar population
which dominates the light in the V band.
The time evolution of these indices reflects that of $Z_V$ because
both indices lie in the V filter, but their metallicity dependence is different,
as already shown by Worthey (1994). He calculates a Z sensitive parameter, which is
the ratio of the percentage change in age to the percentage change in Z. This
parameter is higher for the iron indices than for the magnesium indices, i.e. the
iron indices as defined by Gorgas are more Z sensitive.
The iron index increases faster than the 
$Mg_2$ index in early galaxy types.
It should be noted that the rate of time evolution is different for different indices
and evolutionary epochs, as well as for broad band colors.
For simple stellar populations, such
as star clusters, Fritze -v. Alvensleben \& Burkert
(1995) have shown that this effect allows to better disentangle age and
metallicity, particular at early stages. For a stellar metallicity confirmed by
spectroscopic observations, ages derived from the observed Fe5270 index agree
well with ages derived from the broad band color (V-I).

It is clear that for a composite stellar system, e.g. a galaxy, with its different 
metallicity and age subpopulations, disentangling both effects becomes more 
difficult.

The theoretical $Mg_2 $ and iron indices in our one - zone model galaxies are 
expected to be a
lower limit to observed values because 
these observations typically refer to the central parts of galaxies.
Moreover, in ellipticals the $Mg_2$ index is higher in mass rich galaxies, as 
expressed in the 
well known luminosity - metallicity relation, while our models refer to a mean
elliptical galaxy luminosity of $M_B \sim 20.5$ at an age of 15 Gyr.
To reduce this shortcoming of the one - zone model we 
compare it with indices determined from large aperture spectra of galaxies.
Another way of testing our model is to 
simulate the central region of early type galaxies with SSPs of various 
metallicities.

\titleb{Comparison with observations of Mg$_2$ and [MgFe]}

The Mg$_2$ observations on globular clusters from Brodie \& Huchra (1990)
are well reproduced
with our Single-Burst - models (cf. Einsel et al. 1995). 
These observed Mg$_2$ values of globular clusters in the regime of 
subsolar metallicities are in the range of 0.05 $\le Mg_2$ [mag] $\le $ 0.20. 
Calibrations for $Mg_2$ vs. stellar [Fe/H]$_*$ are given by Burstein et al. 
(1984) as derived from observations
of globular clusters and by Buzzoni et al. (1992) as derived from their evolutionary
synthesis models.
The work of Matteucci \& collaborators (1987 ff) has shown that the iron abundance
evolution should not be expected to directly correspond with magnesium because of SNI 
contributions.
Instead of the usual $Mg_2$ vs. [Fe/H]$_*$ relation, a calibration between the mean
luminosity weighted stellar metallicity and various indices might be more useful
because both can be observed. 
In our notation, a calibration of Mg$_2$ vs [Fe/H]$_*$ for 
elliptical galaxies could be expressed as 
a relation between Mg$_2$ and Z$_V$. From the monotonic
increase of Mg$_2$ with time (Fig. 8) and the enrichment and 
subsequent dilution behaviour of Z$_V$(t) (Fig. 4), it is clear that 
no unique relation between  Mg$_2$ and Z$_V$ is obtained for our composite 
metallicity elliptical galaxy model.
Observations of stellar metallicities in different 
wavelength regions might even provide information about the SF history
of the galaxy.

Fig. 10 shows the time evolution of the stellar index [MgFe]. In this
figure we compare the SSPs for three metallicities with the chemically consistent
model of an elliptical galaxy. As expected from Figs. 8 and 9 
the [MgFe] evolution of our composite elliptical galaxy model remains below that
of a half solar SSP model
over a Hubble time. 
Observations of [MgFe] by Gonzales (1993) 
of the central regions of elliptical galaxies, i.e. within Re/8 and Re/2,
where Re is the effective radius (see also Bressan et al. 1995), lie between
0.4$\le $ log[MgFe]$\le $ 0.6, with the bulk of the Re/2 data at $\sim $0.53.
The
evolution of [MgFe] from the SSPs with Z=$10^{-2}$ and Z=2Z$_{\odot}$
fit these data well, while as 
expected, our global elliptical model with [MgFe] $\sim $ 0.48 
only gives a lower limit to the observed central [MgFe]. 

\titleb{Relations between indices and ISM abundances}

\begfig 6 cm
\vskip -6.5cm
\epsfysize=6cm
{\epsffile{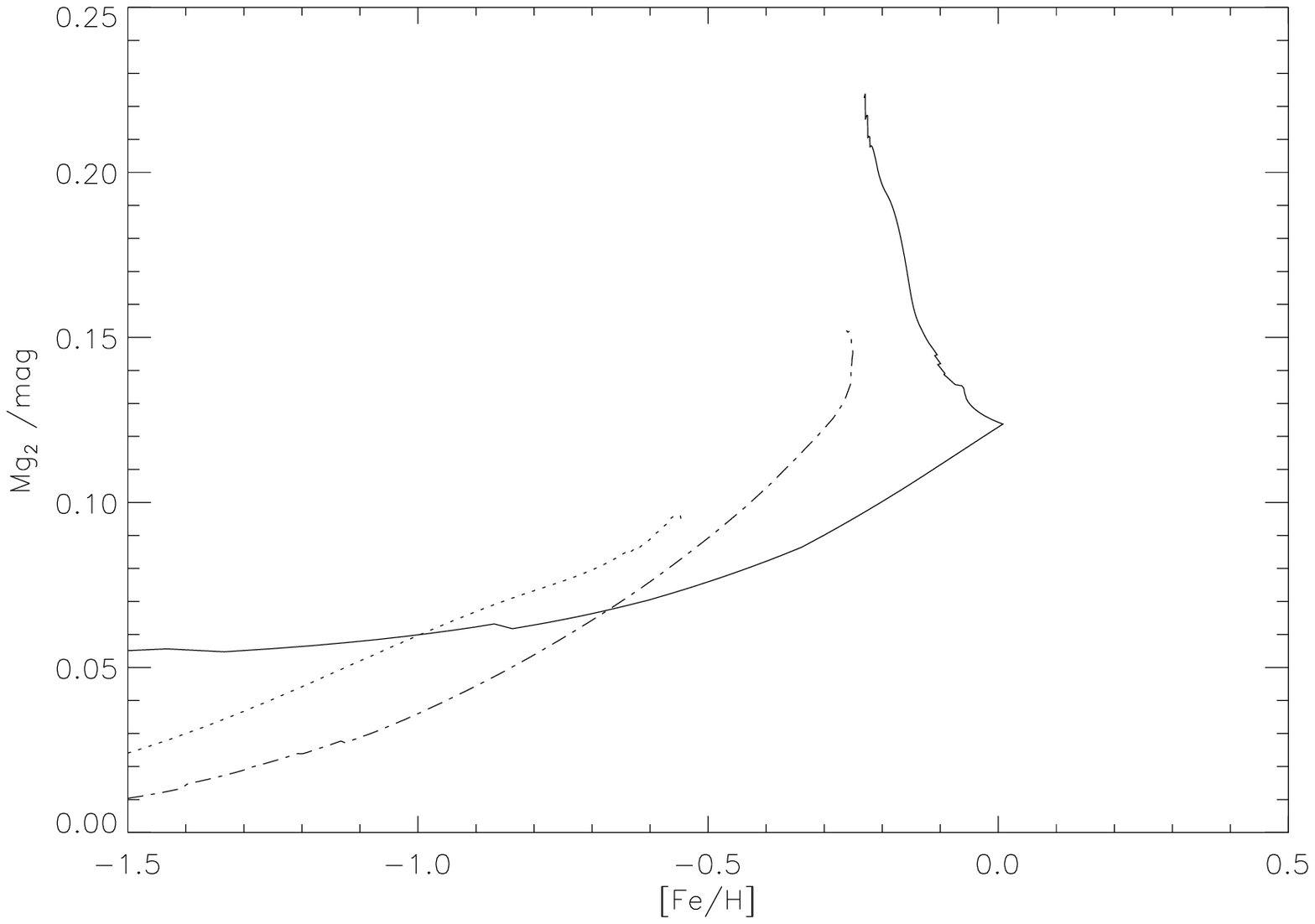}}
\figure{11}{Stellar index Mg$_2$ versus [Fe/H]$_{ISM}$. The solid curve shows the time 
evolution for E - galaxies, the dot-dashed that for Sb - galaxies and
the dotted that for Sd - galaxies.}
\endfig

\begfig 6 cm
\vskip -6.5cm
\epsfysize=6cm
{\epsffile{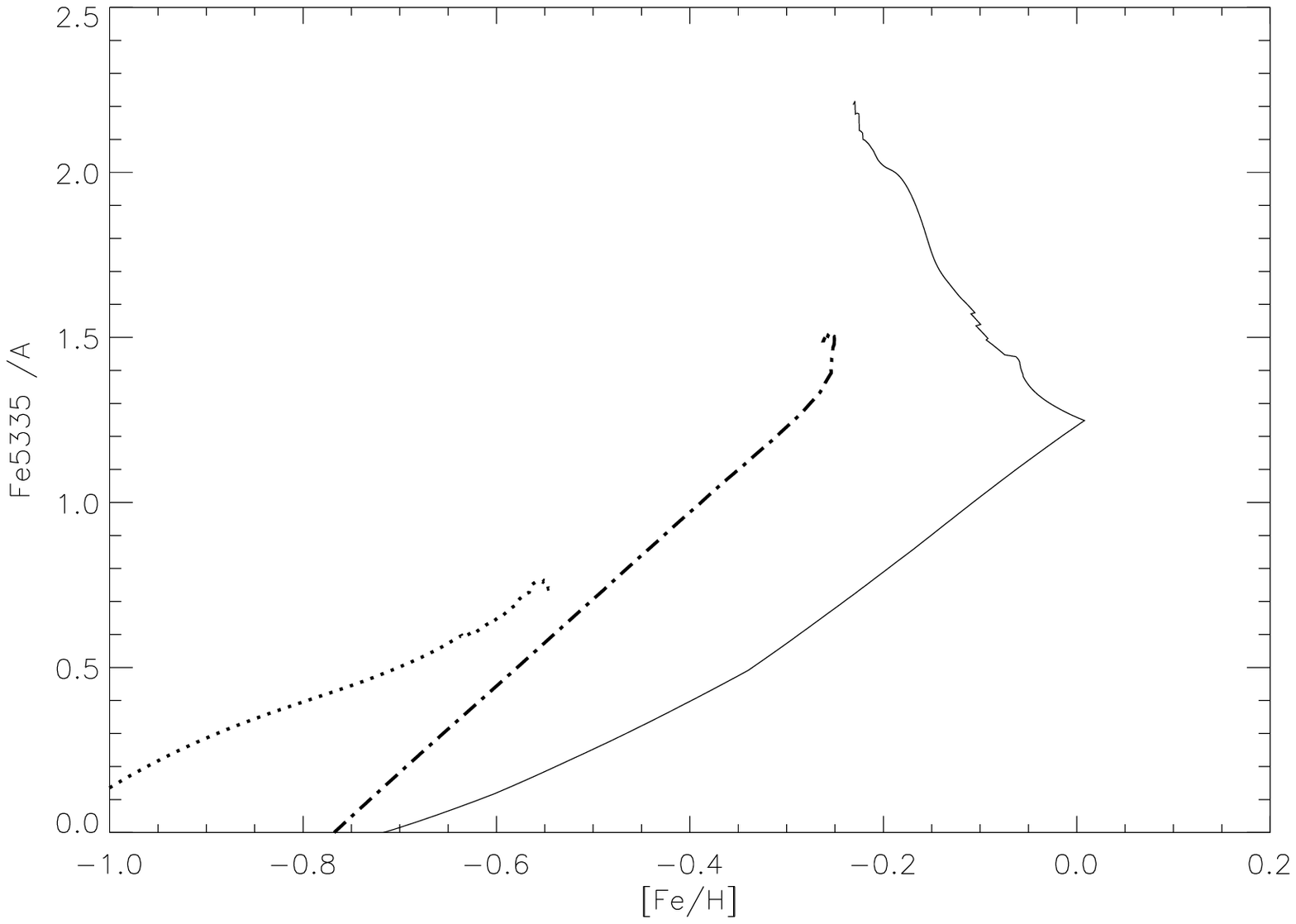}}
\figure{12}{Time evolution of stellar index Fe5335 vs. [Fe/H]$_{ISM}$ for ellipticals
(solid), Sb - galaxies (dot-dashed) and Sd - galaxies (dotted).}
\endfig

It is interesting to find a relation between stellar indices and the ISM metallicity
despite the fact that in the past, stellar indices have generally been measured from
elliptical galaxies while ISM abundances were determined from HII - region observations
in spiral galaxies.

Cold molecular gas has now been 
found in several ellipticals (cf. e.g. Lees et al. 1991) and
abundance determinations from sub-mm lines may be possible in the near future.
X-ray observations of the hot coronae of massive ellipticals has already 
provided abundances
for the diluted hot gas, which is
possibly being blown out of the galaxies (Arimoto 1995). To
determine the origin of this gas, a comparison with model ISM properties may be
useful, though in this case a generalization of our simple closed-box model would be 
required.

QSO absorption line observations and optical identification of absorbers point to
large gaseous halos containing sufficient amounts of magnesium and carbon to overcome
the lower equivalent width thresholds of these observations around early and possibly
even late type galaxies (see Dickinson 1995, Tytler 1995 for a recent review). 
These observations 
will soon offer a unique opportunity to model at the same time the spectral signature of
the galaxy, including stellar absorption features {\bf and} its ISM abundances.

Here, we will use our chemically consistent photometric evolutionary model, together with
a detailed description  of the iron enrichment in the ISM 
to derive relations
of stellar $Mg_2$ and Fe5335 with [Fe/H]$_{ISM}$ as a function of star formation 
history. We can then study how these 
relations evolve as a function of galaxy age.
We thus obtain theoretical relations between metal
indices and gas metallicities, by modelling the enrichment of 
$^{56}Fe$. 

The difference between the evolution of $^{56}Fe$ and $^{24}Mg$ is that the dilution
of $^{56}Fe$ is not as strong as that of $^{24}Mg$ due to the continuing iron -
enrichment from SNeI. 
The evolution of the global gas metallicity Z is similar
to that of $^{16}O$ and $^{24}Mg$ because 
these typical SNII products make up the bulk of the heavy elements.

The time evolution of
the relation between stellar indices $Mg_2$, $Fe5335$ and gas metallicity
given in terms of [Fe/H]$_{ISM}$ is presented in Figs 11 and 12 for E, Sb and Sd 
galaxies. While the values of both the stellar indices and [Fe/H]$_{ISM}$ in 
spirals increase 
continuously with time (Fig. 11, 12), the relation for ellipticals is not unique.
This is 
because of the maximum and later dilution in the ISM of [Fe/H]$_{ISM}$ and the continuous
increase of Mg$_2$ (Fig. 8). While the lower branch of the evolution for ellipticals
belongs to very early phases ($\le$ 3 Gyr), the upper one, which is roughly 
perpendicular to the relation for spirals, gives the evolution after the ISM
abundance maximum.

These calibrations between stellar metal indices and [Fe/H]$_{ISM}$ are obviously
a first approximation and have to be extended to also include dynamical effects, such
as e.g. mass loss through SN-driven galactic winds or the formation of abundance
gradients. 
Dropping our simplifying 
one - zone, closed box assumption would result in a metallicity gradient 
and imply metallicities that will probably be 
lower in the outer part and higher in the central regions as 
compared to our overall metallicity.
In the central regions the
metallicity may even increase to a few times solar, so that additional stellar tracks
of higher metallicity become necessary.
All this, however, is beyond the scope of the present paper.

\titlea{Summary}

Using our chemically consistent evol\-utio\-na\-ry  
method, 
which follows the enrichment of metallicity
in the ISM and in successive generations of stars,
we are able to reproduce the observed colors for 
different spectral types of galaxies. 
We obtain good agreement of our
synthetic galaxy spectra at ages of 11, 12 and 8 Gyr for E -, Sb - and Sd -
model galaxies with
Kennicutt's  observed templates. Our model ISM abundances also agree
with Zaritzky's observations of various spiral types.

We analyse the evolution of and relations between different
measurements of the metallicity: abundances of heavy elements in the ISM,  
luminosity weighted stellar metallicities in different wavelength
bands, and various stellar absorption
indices. 
We study the influence of the star formation history on luminosity weighted 
stellar metallicities and indices.

Important differences are found between
luminosity weighted stellar metallicities at 
different wavelengths. For example in ellipticals,
the stellar metallicity seen in the K - band is predicted to be higher by
a factor of two than that seen in U.
Going from the U to the K band the trend for luminosity weighted stellar 
metallicities in ellipticals is just the inverse to that for spirals, i.e. in
spiral type galaxies the luminosity weighted stellar metallicity is higher in U 
than in V and K.
For the Sb model the discrepancy between stellar and ISM abundances is larger
than in the E or Sd model.
The ISM metallicity of our Sb model is a factor of two higher than the 
luminosity weighted stellar metallicity in any wavelength region.

We analyse the luminosity contributions to various filters from stellar 
subpopulations of different metallicities for different spectral types 
of galaxies.
While the metallicity distribution
of stars contributing light to a passband reaches a maximum at $Z\sim 0.5Z_{\odot}$
and  then declines towards
the highest metallicities at all wavelength for ellipticals, it shows a monotonic,
though overall weaker, increase in spirals. In addition, 
the maximum of the
metallicity distribution in ellipticals depends on wavelength. 
At higher metallicity it is at 
shorter wavelength.
Therefore, knowing the metallicity distribution allows one to constrain the age 
distribution for a known SF history in composite stellar populations.

We show for spectral types from E to Sd how the time evolution of the stellar
metallicity indices Mg$_2$, Fe5335 and [MgFe] depend on the SF history. 
These indices increase
continuously with time, with the iron index growing the fastest, which
reflects the fact that it
is more sensitive to metal enrichment than Mg$_2$.
Although the stellar metal indices Mg$_2$ and Fe5335 lie in the V band, there is no 
unique relation with the luminosity weighted stellar metallicity
Z$_V$, because Z$_V$ 
reaches a maximum and is subsequently diluted.

We also calculate simple stellar populations for 
various metallicities and compare their [MgFe] time evolution with that of the
chemically consistent model and observations from the central regions of ellipticals
(Gonzales 1993). Our simple stellar populations with Z=2Z$_{\odot }$ and 
Z=0.5Z$_{\odot }$ fit well the observed central [MgFe], while our global E model 
as expected gives a lower limit. 

Our models also allow one to study stellar features together with the evolution of ISM
abundances, e.g. [Fe/H]$_{ISM}$. We present theoretical relations between Mg$_2$ and
Fe5335 vs [Fe/H]$_{ISM}$ that will be useful for a consistent chemical and spectral
interpretation of optically identified QSO absorbers.

While in our elliptical model the enrichment of the 
ISM does not directly correlate in a unique way throughout an entire
Hubble time with that of 
the stellar population, we find a linear relation between stellar and ISM metallicity
in spiral galaxy models.

\vskip 1 cm

{\sl Acknowledgements.}
We are grateful to Robert C. Kennicutt, Jr. for sending us 
his atlas of template galaxy spectra.
We are also indebted to Gustavo Bruzual for providing us Kurucz's 
stellar atmosphere spectra. 
We thank our referee, Danielle Alloin, for very useful comments.
C. S. M. and U. F.- v.A. acknowledge
financial support from the Deutsche Forschungsgemeinschaft under
grants Fr 325/36-1 and Fr 916/2-1.
The computations were carried out on a HP - Workstationcluster in the
Universit\"atssternwarte.


\begref{References}

\ref Arimoto, N., Yoshii, Y. 1986, A\&A, {\bf 164}, 260

\ref Arimoto, N., Yoshii, Y. 1987, A\&A, {\bf 173}, 23

\ref Arimoto, N. 1995, in {\sl From Stars to Galaxies}, International Conference, 
in press

\ref Bahcall, J.N., Flynn, C., Gould, A. 1992, ApJ, {\bf 389}, 234

\ref Bessel, M.S., Brett, J.M. 1988, PASP, {\bf 100}, 1134 

\ref Bica, E., Alloin, D. 1986, A\&A, {\bf 162}, 21 

\ref Bica, E., Alloin, D. 1987, A\&A SS, {\bf 70}, 281 

\ref Bica, E. 1988, A\&A, {\bf 195}, 76

\ref Bica, E., Arimoto, N., Alloin, D. 1988, A\&A, {\bf 202}, 8

\ref Bressan, A., Chiosi, C., Tantalo, R. 1995, A\&A, in press

\ref Brodie, J.P., Huchra, J.P. 1990, ApJ {\bf 362}, 503

\ref Bruzual A., G. 1993, in {\sl The enviroment and Evolution of Galaxies,} eds.
Shull, J.M., Thronson, H.A., Kluwer Academic Puplishers, p.91

\ref Bruzual A., G., Charlot, S. 1993, ApJ {\bf 405}, 538

\ref Burstein, D., Faber, S.M., Gaskell, C.M., Krumm, N. 1984, ApJ {\bf 287}, 586

\ref Buzzoni, A., Gariboldi, G., Mantegazza, L. 1992, AJ {\bf 103}, 1814

\ref de Vaucouleurs, G., de Vaucouleurs, A., Corwin, H.G., Buta, R.J., Paturel, G.,
Fouqu\'e, P. 1991, in {\sl Third Reference Catalogue of Bright Galaxies,} Springer
Verlag

\ref Dickinson, M. 1995, in {\sl New Light on Galaxy Evolution}, IAU Symposium 171,
in press

\ref Einsel, Ch. 1992, Diploma Thesis, University of G\"ot\-tin\-gen

\ref Einsel, Ch., Fritze - v. Alvensleben, U., Kr\"uger, H., Fricke, K.J. 1995,
 A\&A {\bf 296}, 347 

\ref Edwardsson, B., Andersen, J., Gustafsson, G., Lambert, D.L., Nissen, P.E.,
Tomkin, J. 1993, A\&A {\bf 275}, 101

\ref Faber, S.M., Tracer, S.C., Gonzalez, J.J., Worthey, G. 1995, in
{\sl Stellar Populations,} IAU Symp. \# 164, eds. van der Kruit, P.C., Gilmore, G.,
Kluwer, Dordrecht, p.249

\ref Fritze - v. Alvensleben, U. 1989, PhD Thesis, University of G\"ottingen

\ref Fritze - v. Alvensleben, U., Gerhard, O.E. 1994, A\&A {\bf 285}, 751

\ref Fritze - v. Alvensleben, U., Burkert, A. 1995, A\&A {\bf 300}, 58

\ref Frogel, J.A. 1985, ApJ {\bf 298}, 528

\ref Gonzales, J.J. 1993, Ph.D. Thesis, Univ. California, Santa Cruz

\ref Gorgas, J., Faber, S.M., Burstein, D., Gonzalez, J.J., Courteau, S., Prosser, C. 
1993, ApJS {\bf 86}, 153

\ref Green, E.M., Demarque, P., King, C.R. 1987, in {\sl The Revised Yale Isochrones
and Luminosity Functions,} Yale Univ. Obs.

\ref Guiderdoni, B., Rocca - Volmerange, B. 1987, A\&A {\bf 186}, 1

\ref Gunn, J.E., Stryker, L.L. 1983, A\&A SS {\bf 52}, 121

\ref Kennicutt, R.C. 1992, ApJ {\bf 388}, 310

\ref Kilian, J., Montenbruck, O., Nissen, P.E. 1994, A\&A {\bf 284}, 437

\ref Kr\"uger, H., Fritze - v. Alvensleben, U., Loose H.-H. 1995, A\&A {\bf 303}, 41

\ref Kurucz, R.L. 1992, in {\sl The Stellar Populations of Galaxies,} IAU Symp. \# 149,
eds. Barbury, B., Renzini, A., Kluwer, Dordrecht, p.225

\ref Lan\c con, A., Rocca - Volmerange, B. 1992, A\&A {\bf 96},593

\ref Lees, J.F., Knapp, G.R., Rupen, M.P., Phillips, T.G. 1991, ApJ {\bf 379}, 177

\ref Matteucci, F., Tornamb\`e, A. 1987, A\&A {\bf 185}, 51

\ref Matteucci, F., Ferrini, F., Pardi, C., Penco, U. 1991, in {\sl Chemical and 
Dynamical Evolution of Galaxies,} ETS, eds. Ferrini, F., p.577

\ref Matteucci, F. 1994, A\&A {\bf 288}, 57

\ref M\"oller, C.S. 1995, Diploma Thesis, University of G\"ot\-tin\-gen

\ref M\"oller, C.S., Fritze - v. Alvensleben, U., Fricke, K.J. 1995, in 
{\sl Stellar Populations,} IAU Symp. \# 164, eds. van der Kruit, P.C., Gilmore, G.,
Kluwer, Dordrecht, p.426

\ref Nissen, P.E. 1995, in
{\sl Stellar Populations,} IAU Symp. \# 164, eds. van der Kruit, P.C., Gilmore, G.,
Kluwer, Dordrecht, p.109

\ref O'Connell, R.W. 1976, ApJ {\bf 206}, 370

\ref Sandage, A. 1986, A\&A {\bf 161}, 89

\ref Scalo, J.M. 1986, Fundam. Cosm. Phys. {\bf 11}, 1

\ref Taylor, B.J., Johnson, S.B., Joner, M.D. 1987, AJ {\bf 93}, 1253			

\ref Terndrup, D.M., Frogel, J.A., Whitford, A.E. 1991, ApJ {\bf 378}, 742

\ref Tinsley, B.M. 1972, A\&A {\bf 20}, 383

\ref Tytler, D. 1995, in {\sl New Light on Galaxy Evolution}, IAU Symposium 171,
in press

\ref Worthey, G. 1994, ApJS {\bf 95}, 107

\ref Worthey, G., Faber, S.M., Gonzales, J.J., Burstein, D. 1994, ApJS {\bf 94}, 687

\ref Wu {\sl et al.} (eds.) 1983, IUE Ultrav. Spektral Atlas, NASA No. 22

\ref Zaritsky, D., Kennicutt, R.C., Huchra, J.P. 1994, ApJ {\bf 420}, 87
 
\endref

\end